\documentclass[twocolumn,trackchanges]{aastex6}
\RequirePackage{lineno}
\usepackage{amsmath}
\usepackage{amssymb}
\usepackage{amsthm}
\usepackage{natbib,aasdefs,url,bm}
\usepackage{array}
\usepackage{float}
\usepackage{graphicx}
\usepackage{subfigure}
\usepackage{color}

\newcounter{ichi}
\newcounter{ni}
\newcounter{san}
\newcounter{yon}
\def\be{\begin{equation}}
\def\ee{\end{equation}}
\def\ba{\begin{eqnarray}}
\def\ea{\end{eqnarray}}

\def\aap{\ {A\&A}\ }

\def\apjl{\ {ApJL}\ }
\def\apjs{\ {ApJS}\ }

\shorttitle{Multi-messenger High-energy Emission Mechanism from SS 433 Jets}
\shortauthors{Kimura, Murase, \& M\'{e}sz\'{a}ros}

\begin{document}

\title{Deciphering the Origin of the GeV--TeV Gamma-ray Emission from SS 433}

\author{Shigeo S. Kimura\altaffilmark{1,2}, Kohta Murase\altaffilmark{3,4,5,6}, and Peter M\'{e}sz\'{a}ros\altaffilmark{3,4,5}}
\altaffiltext{1}{Frontier Research Institute for Interdisciplinary Sciences, Tohoku University, Sendai 980-8578, Japan}
\altaffiltext{2}{Astronomical Institute, Tohoku University, Sendai 980-8578, Japan}
\altaffiltext{3}{Department of Physics, The Pennsylvania State University, University Park, Pennsylvania 16802, USA}
\altaffiltext{4}{Department of Astronomy \& Astrophysics, The Pennsylvania State University, University Park, Pennsylvania 16802, USA}
\altaffiltext{5}{Center for Multimessenger Astrophysics, Institute for Gravitation and the Cosmos, The Pennsylvania State University, University Park, Pennsylvania 16802, USA}
\altaffiltext{6}{Center for Gravitational Physics, Yukawa Institute for Theoretical Physics, Kyoto, Kyoto 606-8502, Japan}


\begin{abstract}
We investigate hadronic and leptonic scenarios for the GeV--TeV gamma-ray emission from jets of the microquasar SS 433. 
The emission region of the TeV photons coincides with the X-ray knots, where electrons are efficiently accelerated. 
On the other hand, the optical high-density filaments are also located close to the X-ray knots, which may support a hadronic scenario. 
We calculate multi-wavelength photon spectra of the extended jet region by solving the transport equations for the electrons and protons. We find that both hadronic and leptonic models can account for the observational data, including the latest {\it Fermi} LAT result. 
The hadronic scenarios predict higher-energy photons than the leptonic scenarios, and future observations such as with the Cherenkov Telescope Array (CTA), the Large High-Altitude Air Shower Observatory (LHAASO), and the Southern Wide-field Gamma-ray Observatory (SWGO) may distinguish between these scenarios and unravel the emission mechanism of GeV--TeV gamma-rays. Based on our hadronic scenario, the analogy between microquasars and radio galaxies implies that the X-ray knot region of the radio-galaxy jets may accelerate heavy nuclei up to 
ultrahigh energies.
\end{abstract}


\section{Introduction}
SS 433 is a micro-quasar that powers bi-polar precessing jets from the central compact object \citep[see][for a review]{2004ASPRv..12....1F}. The central object is expected to accrete the material at a super-Eddington rate \citep{1981VA.....25...95V}. The jets are interacting with interstellar matter at tens of parsecs, creating a bright radio nebula \citep{1998AJ....116.1842D} and extended X-ray lobes \citep{1980Natur.287..806S,1983ApJ...273..688W,1994PASJ...46L.109Y}. The X-ray lobes have bright knots in both the eastern and western jets \citep{1997ApJ...483..868S,1999ApJ...512..784S}. Recently, the High Altitude Water Cherenkov collaboration (HAWC) reported 20 TeV gamma-rays associated with the X-ray knots \citep{2018Natur.562...82A}, which indicates the existence of particles of at least a few hundreds of TeV. 

Some groups have searched for the high-energy gamma-rays from SS 433 using {\it Fermi} Large Area Telescope (LAT), but the results differ from each other. \citet{2015ApJ...807L...8B} discovered a gamma-ray source at a position consistent with SS 433, but not aligned with the extended jet. \citet{2019ApJ...872...25X} reported detection of GeV gamma-rays with a soft spectrum from the western knot but non-detection from the eastern knot. \citet{2019MNRAS.485.2970R} discovered a periodic gamma-ray emission from the SS 433 region, and argued that the emission comes from the central object. By contrast, \citet{2019A&A...626A.113S} found that the GeV gamma-ray emission region is larger than the TeV gamma-ray emission region, arguing that the GeV source likely originates from W50, a radio nebula surrounding SS 433, rather than the SS 433 knots. 

Recently, \citet{2020ApJ...889L...5F} performed a joint analysis of {\it Fermi} LAT and HAWC data, and concluded that the GeV gamma-ray data may be smoothly connected to the TeV range with a photon index $\Gamma\sim2.1$. They found that the previous {\it Fermi} LAT analyses were affected by nearby sources, J1913.3+0515 in the {\it Fermi} LAT 8-year point source catalog and J1907.9 + 0602 in the 4FGL catalog \citep{2019arXiv190210045T}. Using the different point-source catalogs and response functions leads to various conclusions. 

The gamma-ray emission region coincides with the X-ray knots \citep{1997ApJ...483..868S}. It is widely believed that high-energy electrons accelerated at the knots emit X-rays by the synchrotron mechanism. Thus, most previous works focus on the leptonic scenario for the TeV gamma-ray emission mechanism \citep{2018Natur.562...82A,2019ApJ...872...25X,2020ApJ...889..146S,2020ApJ...889L...5F}.  However, hadronic emission could provide a dominant contribution for the observed gamma-rays \citep{2019APh...109...25R}. Optical filaments exist within the angular uncertainty of the gamma-ray signals in the eastern lobe \citep{1980MNRAS.192..731Z,1983MNRAS.205..471K,2007MNRAS.381..308B}. The  particle number density in the filaments is much higher than in the ambient medium, which motivates us to investigate a hadronic scenario more carefully. 

In this paper, we examine both scenarios using the multi-wavelength data, including the latest GeV data by {\it Fermi} LAT, and discuss the scenario feasibility and tests by future observations. We focus on the eastern lobe. In the western lobe, it is unclear whether dense filaments exist close to the gamma-ray emission region or not, and we avoid discussion of hadronic scenario there.
In Section~\ref{sec:model}, we construct a steady-state one-zone model, and describe the model parameters obtained from multi-wavelength observations. Our calculation results are shown in Section~\ref{sec:results}, and the analogy to large scale jets in radio galaxies is discussed in Section~\ref{sec:RG}. We discuss the implications  in Section~\ref{sec:discussion} and summarize our results in Section~\ref{sec:summary}.
The notation of $Q_X=Q/10^X$ in cgs unit is used unless otherwise noted.

\section{Models}\label{sec:model}

\begin{figure}
    \centering
    \includegraphics[width=\linewidth]{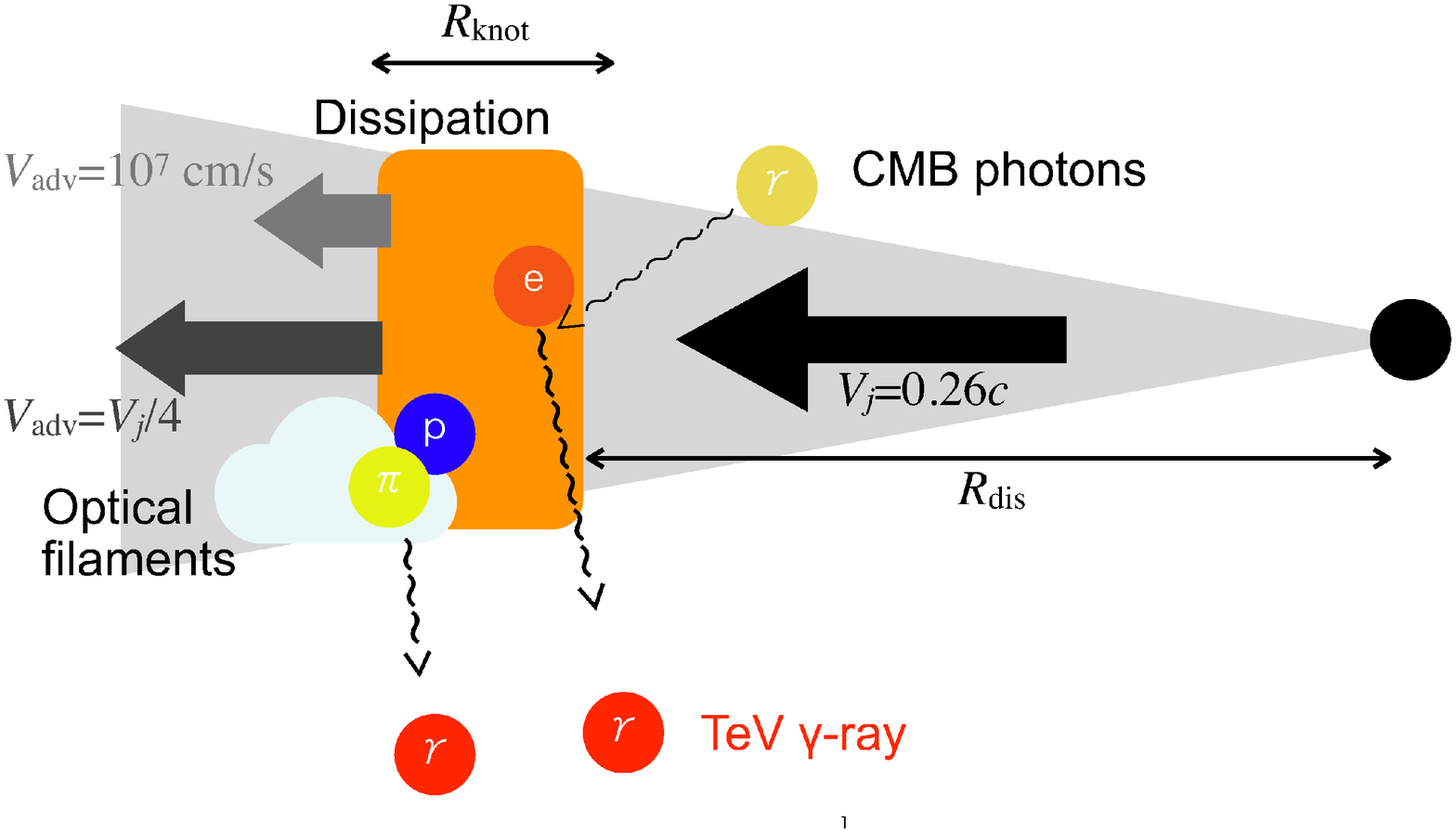}
    \caption{Schematic picture of our models. The jets dissipate their kinetic energy at a dissipation radius, $R_{\rm dis}$, which accelerates non-thermal particles. The non-thermal protons interact with ambient matter including the dense optical filaments, producing gamma-rays through pion decay. The non-thermal electrons emit gamma-rays by up-scattering the CMB photons. We write the size of the emission region as $R_{\rm knot}$. We consider 4 scenarios: combinations of hadronic-dominated/leptonic-dominated and fast (dark-grey)/slow (light-grey) advection velocity (see Table \ref{tab:models}).
    }
    \label{fig:schematic}
\end{figure}

\subsection{Formulation}
We assume that the jets of kinetic luminosity $L_j$ dissipate some of their energy at the X-ray knot, resulting in acceleration of non-thermal particles (see Figure \ref{fig:schematic} for schematic picture). To obtain the particle spectra at the X-ray knot, we solve the steady state transport equation for non-thermal particles of species $i$: 
\begin{equation}
 \frac{d}{dE_i}\left(-\frac{E_i}{t_{i,\rm cool}}N_{E_i}\right)
  = - \frac{N_{E_i}}{t_{\rm esc}}+ \dot N_{E_i} ,
\end{equation}
where $E_i$ is the particle energy ($i=$e or p), $N_{E_i}$ is the total number spectrum, $t_{i,\rm cool}$ is the cooling time, $t_{\rm esc}$ is the escape time, and $\dot N_{E_i}$ is the injection term.
 This equation has an analytic solution (see Appendix C in \citealt{2009herb.book.....D}): 
 \begin{equation}
  N_{E_i}=\frac{t_{i,\rm cool}}{E_i}\int_{E_i}^{\infty}dE'_i \dot N_{E'_i}\exp\left(-\int_{E_i}^{E'_i}\frac{t_{i,\rm cool}}{t_{\rm esc}}d\mathcal E_i\right).
 \end{equation}
We numerically integrate this equation to obtain the proton and electron spectra.
We consider the diffusive shock  acceleration mechanism at the knot and set the injection term to be a power-law form with an exponential cutoff:
\begin{equation}
\dot N_{E_i}=\dot N_{i,\rm nor}\left(\frac{E_i}{E_{i,\rm cut}}\right)^{-p_{\rm inj}}\exp(-\frac{E_i}{E_{i,\rm cut}}),
\end{equation}
where $\dot N_{i,\rm nor}$ is the normalization factor, $p_{\rm inj}$ is the power-law index, and $E_{i,\rm cut}$ is the cutoff energy determined by the balance between acceleration and loss timescales, $t_{\rm loss}^{-1}=t_{\rm cool}^{-1}+t_{\rm esc}^{-1}$. We normalize the normalization factor so that $\int E_i\dot{N}_{E_i}dE_i=\epsilon_iL_j$ is satisfied, where $\epsilon_i$ is the energy conversion factor. 

We assume the same bulk velocity for the electrons and protons. They should have the same acceleration and diffusion timescales at a given energy.
The diffusive shock acceleration time is given by
\begin{equation}
 t_{\rm acc}\approx\frac{20\eta E_i}{3ce B\beta_j^2},
\end{equation}
where $\eta$ is the acceleration efficiency, $B$ is the magnetic field strength, and $\beta_j$ is the jet velocity. As the escape processes, we consider diffusion and advection, whose timescales are estimated to be
\begin{equation}
 t_{\rm diff}\approx \frac{3eBR_{\rm knot}^2}{2c\eta E_i},
\end{equation}
\begin{equation}
 t_{\rm adv}\approx \frac{R_{\rm knot}}{V_{\rm adv}},
\end{equation}
where $R_{\rm knot}$ is the size of the knot and $V_{\rm adv}$ is the advection velocity at the knot.
 Assuming a spherical geometry of the emission region, the adiabatic cooling timescale is expressed as 
\begin{equation}
 t_{\rm adi}\approx \frac{R_{\rm dis}}{V_{\rm adv}},
\end{equation}
where $R_{\rm dis}$ is the distance of the dissipation region from the central object. Note that if the jet geometry is cylindrical, one can ignore the adiabatic cooling \citep{2020ApJ...889..146S}.

For the electron radiation processes, we consider synchrotron and inverse Compton scattering (IC). The synchrotron timescale for the species $i$ is represented as
\begin{equation}
 t_{i,\rm syn}\approx \frac{6\pi m_e^2 c^3}{\sigma_TB^2E_i}\left(\frac{m_i}{m_e}\right)^2,
\end{equation}
where $m_i$ is the mass of the particle $i$ and $\sigma_T$ is the Thomson cross section.  We use a fitting formula (Equation [18]--[20]) in \citet{2008ApJ...686..181F} to calculate the synchrotron spectrum. 
The IC cooling rate is estimated using Equation (2.56) in \citet{1970RvMP...42..237B}, and the IC spectrum is calculated by Equation (2.48) in  \citet{1970RvMP...42..237B}.
We consider only the cosmic-microwave background (CMB) as the target photons, since IC emission using other photon fields is sub-dominant \citep{2020ApJ...889..146S,2020ApJ...889L...5F}.

For the hadronic radiation processes, we consider only the $pp$ inelastic collisions, because other processes are negligibly efficient \citep{2019APh...109...25R}. We should note that for $n_{\rm eff}\lesssim0.01$,  external photon fields by the central star or by the beamed emission from the inner jets may be important.  The $pp$ cooling rate is estimated to be 
\begin{equation}
t_{pp}^{-1}=n_{\rm eff}\sigma_{pp}\kappa_{pp}c,
\end{equation}
where $n_{\rm eff}$ is the effective number density (defined in the following subsection), $\sigma_{pp}$ is the $pp$ inelastic collision cross section given in \citet{2014PhRvD..90l3014K}, and $\kappa_{pp}\approx 0.5$ is the inelasticity for $pp$ interaction. We use the method of \citet{kab06} to calculate the gamma-ray spectrum by $pp$ inelastic collisions. 

 \subsection{Model parameters}
 
Multi-wavelength observations of SS 433 provide useful information to model the high-energy emission from the extended jets. The jet velocity is measured to be $\beta_j\simeq0.26$ at the jet base using both optical \citep{1979Natur.279..701A,2001ApJ...561.1027E} and X-ray data \citep{2002ApJ...564..941M}. The mass loss rate of the jet is estimated to be $\dot M_j\simeq5\times10^{-7}\rm~M_{\odot}~yr^{-1}$ \citep{1983MNRAS.205..471K}, which leads to a kinetic energy of the jet of $L_j\approx\dot M\beta_j^2c^2/2\simeq2\times10^{39}\rm~erg~s^{-1}$. The size and the distance from the central object for the brightest X-ray knot (e2) are 5' and 35', respectively \citep{1997ApJ...483..868S}, which correspond to  $R_{\rm dis}\simeq56$ pc and $R_{\rm knot}\simeq8.1$ pc, with the distance of $d_L=5.5$ kpc. 

Optical observations discovered filamentary structures located close to the X-ray knots \citep{1980MNRAS.192..731Z}, where the number density can be as high as $n\sim10^2\rm~cm^{-3}$ \citep{1983MNRAS.205..471K} and the velocity is estimated to be $V_{\rm adv}\sim10^7\rm~cm~s^{-1}$ \citep{2007MNRAS.381..308B}. On the other hand, \citet{2017A&A...599A..77P} estimate the mean number density in W50 to be $n\sim0.1\rm~cm^{-3}$, and argue that the jet is not significantly decelerated at the X-ray knot. In this case, the bulk velocity of the emission region is likely to be $V_{\rm adv}\approx\beta_jc/4$, where the factor 4 indicates energy dissipation by a strong shock. Since the advection velocity and the target gas density in the X-ray knot are still largely uncertain, we examine two values of the advection velocity: $V_{\rm adv}=\beta_jc/4\simeq1.9\times10^9\rm~cm~s^{-1}$ (scenarios A \& C)  or $V_{\rm adv}=10^7\rm~cm~s^{-1}$ (scenarios B \& D). Even for the low advection velocity cases, we assume a shock velocity of $\beta_j$, because the accelerated electrons cannot emit the observed X-rays with a lower value of the shock velocity (see Section \ref{sec:results}). Regarding the number density, we define the effective number density as $n_{\rm eff}=f_{\rm fil}n_{\rm fil}$, where $n_{\rm fil}\sim100\rm~cm^{-3}$ and $f_{\rm fil}\sim10^{-4}-1$ are the number density and the volume filling factor of the optical filaments, respectively. 
Here we note that the magnetic field strength and the effective number density are treated as independent parameters. Also, because we assume $f_{\rm fil}\ll1$ in our scenarios, we should evaluate the magnetic field strength at the X-ray knot, and the magnetic field strength does not have to scale with the effective density.

\section{Results}\label{sec:results}

\begin{table}
\begin{center}
\caption{Model parameters in our lepto-hadronic scenarios; scenarios A and B
  are hadronic-dominated, while C and D are leptonic-dominated.}\label{tab:models}
Fixed parameters
\begin{tabular}{cccccccc}
\hline
\hline
$\beta_j$ & $L_j$ & $R_{\rm knot}$ & $R_{\rm dis}$ & $\epsilon_p$ & $\eta$ &  $d_L$ \\
 & [$\rm erg~s^{-1}$] & [pc] & [pc] & & & [kpc] \\
\hline
0.26 & $2\times10^{39}$ & 8.1 &  56 & 0.1 & 2 & 5.5 \\
\hline
\hline
\end{tabular}
 
Model parameters.
\begin{tabular}{cccccc}
\hline
\hline
Scenario & $V_{\rm adv}$ & $B$ & $p_{\rm inj}$ & $\epsilon_e$ & $n_{\rm eff}$ \\
 & [$\rm cm~s^{-1}$]  & [$\rm\mu$G] &  &  & [$\rm cm^{-3}$] \\
\hline
A & $1.9\times10^9$ & 32& 2.0 &  $1.0\times10^{-3}$ & 10 \\
B & $1.0\times10^7$ & 36 & 1.6 &  $1.5\times10^{-4}$ & 0.2 \\
C & $1.9\times10^9$ & 13 & 2.1 & $5.0\times10^{-3}$ & 0.01 \\
D & $1.0\times10^7$ & 18 & 1.6 & $2.0\times10^{-4}$ & 0.01\\
\hline
\hline
\end{tabular}
\end{center} 
\end{table}

  \begin{figure*}
   \begin{center}
    \includegraphics[width=\linewidth]{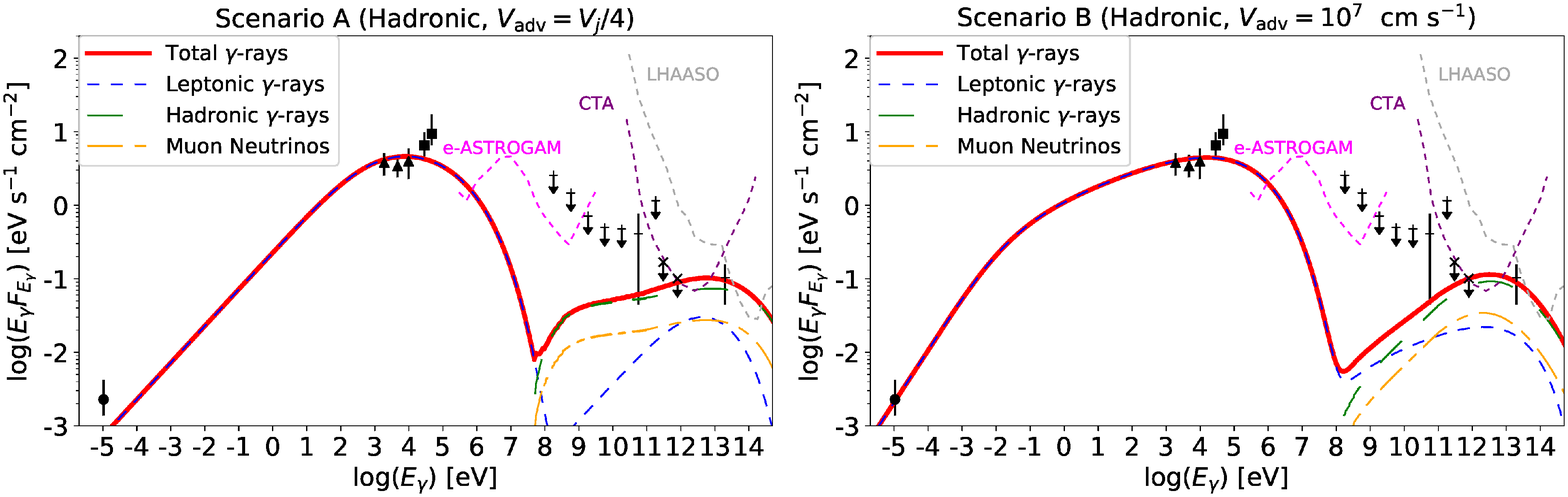}
    \includegraphics[width=\linewidth]{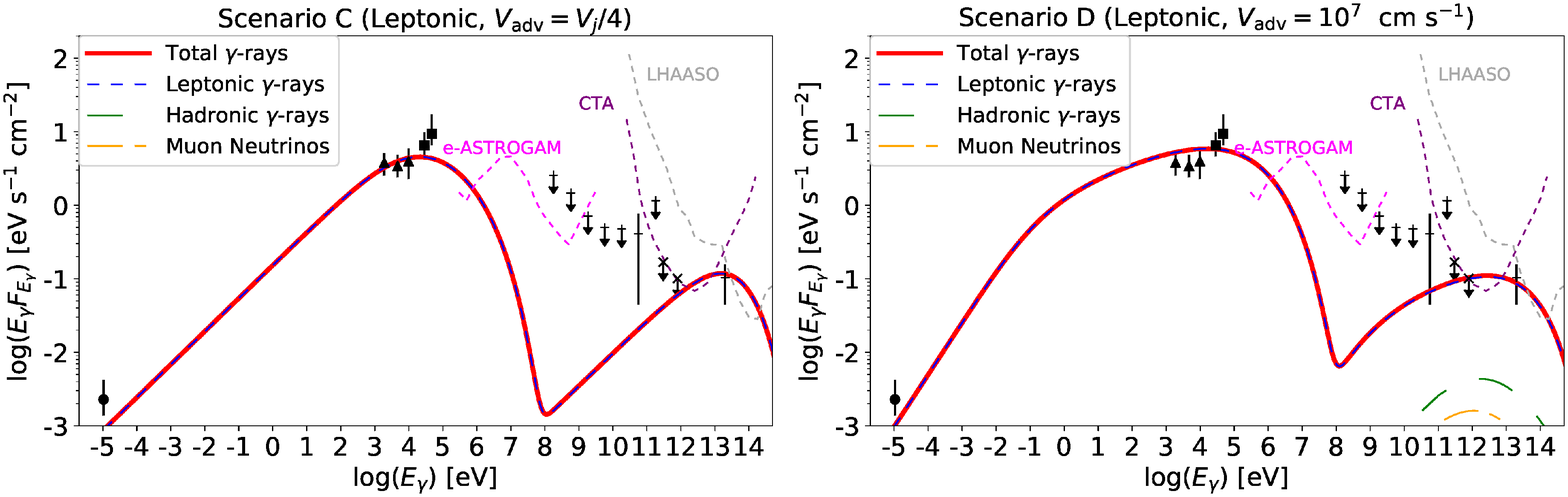}
    \caption{Photon spectra from the extended jets of SS 433 for scenario A (top-left), B (top-right), C (bottom-left), and D (bottom-right). The red-thick-solid, green-thin-long-dashed, and blue-thin-short-dashed lines are total, hadronic, and leptonic components, respectively. The observational data are taken from \citet{1980A&A....84..237G} (circle), \citet{2007A&A...463..611B} (triangles),  \citet{1997ApJ...483..868S} (squares), \citet{2018A&A...612A..14M} (crosses) and \citet{2020ApJ...889L...5F} (pluses). The thin-dotted lines are sensitivity curves for e-ASTROGAM \citep[1 yr;][]{2017ExA....44...25D}, CTA \citep[50 h;][]{2019scta.book.....C}, and LHAASO \citep[1 yr;][]{2019arXiv190502773B}. Scenarios A, B, and D can reproduce the GeV--TeV gamma-ray data, while scenario C cannot reproduce the {\it Fermi} data. The thin-dotted-dashed lines are the muon neutrino spectra (i.e., neutrino spectra per flaver). Also, the CTA sensitivity curve is for a point source. The TeV gamma-ray emission region in SS 433 is extended, which worsens the sensitivity. }
    \label{fig:fit}
   \end{center}
  \end{figure*}

We calculate the photon spectra for various values of $p_{\rm inj}$, $\epsilon_e$, $B$, and $n_{\rm eff}$ to seek the parameter set that matches the data. Since the radio map of W50 does not indicate any clear knot-like structure \citep{1998AJ....116.1842D}, we should regard the radio data as an upper limit.  
We match the data by eye inspection, and do not discuss the goodness of fit because of the observational uncertainty and the limitation of the models.
Figure \ref{fig:fit} shows both the leptonic and hadronic contributions to the photon spectra for our scenarios whose parameter sets are tabulated in Table \ref{tab:models}. For all the scenarios, the electron synchrotron emission is responsible for the X-ray data. The Lorentz factor of electrons emitting the hard X-rays is estimated to be
 \begin{equation}
  \gamma_{e,X}\approx \sqrt{\frac{4\pi m_e c E_\gamma}{h_p e B }}\simeq4.1\times10^8B_{-4.5}^{-1/2}\left(\frac{E_\gamma}{30\rm~keV}\right)^{1/2},\label{eq:gamex}
 \end{equation}
where $h_p$ is the Planck constant. The synchrotron cooling is the dominant loss process in this energy range for all the scenarios. Equating the synchrotron and acceleration timescales, we obtain the maximum Lorentz factor of the electrons:
\begin{equation}
 \gamma_{e,\rm cut}\approx\sqrt{\frac{9\pi e\beta_j^2}{10\sigma_TB\eta}}\simeq2.1\times10^9B_{-4.5}^{-1/2}\eta_0^{-1/2}.\label{eq:Eemax}
\end{equation}
From the condition $\gamma_{e,X}<\gamma_{e,\rm cut}$, we obtain an upper limit for $\eta$:
\begin{equation}
 \eta\approx\frac{9}{40}\frac{h_pe^2\beta_j^2}{\sigma_Tm_ecE_\gamma}\simeq27\left(\frac{E_\gamma}{30\rm~keV}\right)^{-1}.\label{eq:eta}
\end{equation}
Thus, the particle acceleration should be very efficient. The synchrotron cutoff feature should be detected by the proposed MeV satellites, such as e-ASTROGAM \citep{2017ExA....44...25D}, All-sky Medium Energy Gamma-ray Observatory \citep[AMEGO;][]{2017ICRC...35..798M}, or Gamma-Ray and AntiMatter Survey \citep[GRAMS;][]{2020APh...114..107A}, which will provide a better constraint on the value of $\eta$.

The synchrotron-cooling break energies for photons and electrons are respectively estimated to be 
\begin{equation}
 E_{\gamma,\rm br}\approx\frac{h_peB\gamma_{e,\rm br}^2}{4\pi m_ec}\simeq0.70B_{-4.5}^{-3}V_{\rm adv,9.3}^2\rm~keV,\label{eq:Egamsyn}
\end{equation}
\begin{equation}
 \gamma_{e,\rm br}\approx\frac{6\pi m_ecV_{\rm adv}}{\sigma_TB^2R_{\rm knot}}\simeq6.2\times10^7B_{-4.5}^{-2}V_{\rm adv,9.3}.
\end{equation}
The break energy lies between the radio and X-ray data points, and $E_{\gamma,\rm br}$ is lower for a lower value of $V_{\rm adv}$ and a higher value of $B$. A lower value of $E_{\gamma,\rm br}$ increases the radio flux if we fix $p_{\rm inj}$ and X-ray luminosity. To avoid overshooting the radio data, a hard spectral index is required for a lower value of $V_{\rm adv}$. 
For scenarios A and C, $p_{\rm inj}$ is consistent with the prediction by the diffusive shock acceleration theory \citep{Bel78a,BO78a}, whereas scenarios B and D demand a harder spectrum that can be realized by the stochastic acceleration mechanism \citep[e.g.][]{bld06,sp08,kmt15,2020PhRvL.125a1101M}. 

As far as the hadronic components, the hadronic gamma-ray spectra roughly follow the parent proton spectra, which have a break due to the diffusive escape. Setting $t_{\rm diff}=t_{\rm adv}$, the proton break energy is estimated to be: 
\begin{equation}
 E_{p,\rm br}\approx\frac{3eBR_{\rm knot}V_{\rm adv}}{2c\eta}\simeq23B_{-4.5}V_{\rm adv,9.3}\eta_0^{-1}\rm~PeV
\end{equation}
For scenario A, the proton spectrum is a single power-law for $E_p\lesssim0.1$ PeV owing to a higher break energy. This naturally makes a power-law gamma-ray spectrum consistent with the observed data. 
 This feature should be detected by the Cherenkov Telescope Array \citep[CTA;][]{2019scta.book.....C}, the Large High-Altitude Air Shower Observatory \citep[LHAASO;][]{2019arXiv190502773B}, and Southern Wide-field Gamma-ray Observatory \citep[SWGO;][]{2019arXiv190208429A}. The detection of a relatively hard sub-PeV gamma-ray  spectrum is a smoking-gun to distinguish the emission mechanism, because the IC up-scattering of CMB photons cannot produce such a feature due to the Klein-Nishina suppression, as shown in the bottom panels of Figure \ref{fig:fit}.
For scenario B, the diffusive break energy is $E_{p,\rm br}\simeq60$ TeV. This produces a peak at $E_\gamma\sim6$ TeV, and the gamma-ray spectrum is softer above that energy. In this case, we cannot discriminate the emission mechanism using the gamma-ray spectrum.
The proton maximum energy is determined by the diffusive escape in all the scenarios:
\begin{equation}
 E_{p,\rm diff}\approx \frac{3eB\beta_jR_{\rm knot}}{\sqrt{40}\eta}\simeq29B_{-4.5}\eta_0^{-1}\rm~PeV.
\end{equation}
This energy is so high that SS 433 can accelerate protons above PeV energies (see Section \ref{sec:discussion} for the possible effects of PeV protons). 

For leptonic scenarios, the GeV-TeV gamma-rays are attributed to IC up-scattering. In scenario C, the high advection velocity makes the break energy too high to match the observation. This cannot make  a flat spectrum in the GeV-TeV range, thus failing to explain the {\it Fermi} data, as in \cite{2020ApJ...889..146S}. On the other hand, in scenario D, the advection time is comparable to the estimated age of the system (30-100 kyr). This enables us to reproduce the broadband spectrum owing to a lower cooling break energy. The resulting spectrum is similar to that by \citet{2020ApJ...889L...5F}.

In our scenario, the magnetic field is unlikely to be generated by some plasma instabilities. The magnetic field in the downstream is often estimated using the $\epsilon_B$ parameter to be
\begin{equation}
 B=\sqrt{\frac{8\epsilon_BL_j}{R_{\rm knot}^2\beta_jc}}.
\end{equation}
With our choice of $B$, $\epsilon_B$ is estimated to be 0.3, 0.4, 0.05, and 0.1 for scenarios A, B, C, and D, respectively. These values are much higher than the values obtained by PIC simulations of non-relativistic shocks \citep{CS14b,PCS15a} and afterglow lightcurve fittings of gamma-ray bursts \citep{2014ApJ...785...29S}. In the leptonic scenario, the observed flux ratio of X-rays to TeV gamma-rays requires the magnetic field strength of $B\sim15\rm~\mu G$, i.e., $\epsilon_B\sim0.1$ \citep{2019ApJ...872...25X,2020ApJ...889..146S,2019APh...109...25R,2020ApJ...889L...5F}. In order for hadronic scenarios to work, a higher magnetic field strength is necessary, and hence,  $\epsilon_B\gtrsim0.2$ is required. 
Note that magnetic fields in our models are not strong compared to that in the interstellar medium (ISM; $B\sim1-10~\rm \mu G$). Shock compression of the ISM field suffices to achieve the values, although it cannot generate the magnetic field of $B>40\rm~\mu G$ for a typical ISM value, $B_{\rm ISM}\sim10\rm~\mu G$. Hence, the magnetic field should be in the range of $10-40\rm~\mu G$ regardless of the emission mechanism.

\section{Comparison to Jets in Radio Galaxies}\label{sec:RG}

In our scenarios, the synchrotron cooling timescales for X-ray emitting electrons are estimated to be $ t_{\rm syn}\simeq78B_{-4.5}^{-2}\gamma_{e,8.5}^{-1}$ yr. This is shorter than the advection timescale for all the scenarios, $t_{\rm adv}\simeq4.0\times10^2V_{\rm adv,9.3}^{-1}$ yr. This demands in-situ particle acceleration in the extended jet with a low $\eta$. This situation may be similar to some of the extended jets in radio galaxies, where the in-situ electron acceleration is required. In our assumption for the advection velocity, scenarios A and C corresponds to the X-ray knot in radio galaxies because the jets are unlikely to be appreciably decelerated at the knots. On the other hand, scenarios B and D are similar to hotspots in radio galaxies, since the termination shock significantly decelerates the plasma flow and forms the cocoon surrounding the jets. 

M87 and 3C 273 are very bright radio galaxies, and the broadband spectra and velocities of their knots are observed. For M87, the knots in the 10--100 pc scale have soft X-ray spectra without a cutoff feature \citep{2018ApJ...858...27Z}. Their intrinsic velocity is estimated to be $\Gamma\beta\sim0.3-10$ with a possible velocity stratification \citep{2019ApJ...887..147P}. On the other hand, for 3C 273, the X-ray spectra for the knots in the kpc-scale is hard, and non-detection by {\it Fermi} suggests a second electron population \citep{2014ApJ...780L..27M}. Their apparent velocities are consistent with $\beta_{\rm app}\sim0$ \citep{2017Galax...5....8M}. For both objects, X-rays are attributed to the in-situ accelerated electrons, which suggests a low value of $\eta\lesssim300$ based on Equation (\ref{eq:eta}) with $\beta_j\sim0.3$. On the other hand, the jet velocity may be very different in these objects, and hence, the value of $\eta$ should be independent of the jet velocity.

According to \cite{2018ApJ...858...27Z}, the peak frequency of the synchrotron spectrum and the magnetic field strength in X-ray knots and hotspots in radio galaxies are estimated to be $\nu_{\rm pk}\sim10^9-10^{17}$ Hz and $B\sim10-300~\rm\mu$G, respectively. 
The radio, optical, and X-ray spectra for some knots are inconsistent with a single component synchrotron emission. A popular interpretation of the emission mechanism for the X-ray emission from the knots is inverse Compton scattering of CMB photons (IC/CMB model; \citealt{TMS00a,2004ApJ...608..698S,2012ApJ...759...86W}). However, non detection of GeV gamma-rays by {\it Fermi} LAT ruled out an IC/CMB model for several sources \citep[e.g.,][]{2017ApJ...849...95B}. The two-component synchrotron model is favored as an alternative scenario for those sources \citep{2004ApJ...613..151A}, which indicates a low value of $\eta$ as in M87 and 3C 273. The shock velocities at the hot spots or X-ray knots are often assumed to be $0.2c-0.5c$ \citep{2005APh....23...31C}, which is also supported by the radio observations of kpc-scale jets \citep{1997MNRAS.286..425W,2004MNRAS.351..727A,2009MNRAS.398.1989M}. Hence, some X-ray knots in radio galaxies should have a very low $\eta$, which is consistent with our SS 433 models.

If protons are accelerated at the X-ray knots simultaneously, the maximum energy of the protons are estimated to be $E_{p,\rm diff}\sim45B_{-4.5}\eta_0^{-1}\beta_{j,-0.5}R_{j,22.5}$ EeV. The iron nuclei can be accelerated up to 26 times higher energies than protons, and hence, the kpc-scale jets in radio galaxies can accelerate heavy nuclei to ultrahigh-energies \citep[see also][]{Tak90a}. However, reproducing the heavy composition obtained by the Pierre Auger Observatory \citep{Aab:2014kda} is challenging by the standard shock acceleration, and re-acceleration of galactic cosmic rays by jets may be important \citep{Cap15a,KMZ18a}.

We should note that the value of $\eta$ should be much higher at hotspots in radio galaxies and blazar zones. The cutoff frequency in the hotspots are estimated to be below the UV range, $\nu_{\rm cut}\lesssim10^{15}$ Hz, leading to $\eta\gtrsim10^4$ \citep{2016MNRAS.460.3554A,2018ApJ...858...27Z}. Also, the IC/CMB model is still favored for some X-ray knots \citep{2018ApJ...858...27Z}, resulting in $\eta$ similar to those in the hotspots. Fittings of the broadband spectra for blazars require $\eta\gtrsim10^4$ \citep{1996ApJ...463..555I,2016ApJ...828...13I,2017MNRAS.464.4875B}. These may indicate that different particle acceleration mechanisms take place at the various places in the astrophysical jets.

\section{Discussion}\label{sec:discussion}

\subsection{Neutrino detectability}

Hadronic TeV gamma-rays must be accompanied by neutrinos of similar energies and fluxes. The eastern lobe of SS 433 is located at the declination of $\delta=+4.9$ degree, so it is also an interesting target for IceCube and IceCube-Gen2. When the neutrino mixing is assumed, the ratio of pionic gamma-rays to muon neutrinos is approximately $2:1$ when the main neutrino production channel is the inelastic $pp$ reaction \citep[e.g.,][and references therein]{mal13}.
Indeed, in scenarios A and B, the predicted neutrino fluxes are $\sim3\times10^{-11}\rm~GeV~cm^{-2}~s^{-1}$ in the $10-100$~TeV range, which is lower than IceCube's ten-year sensitivity of $\sim3\times10^{-10}\rm~GeV~cm^{-2}~s^{-1}$ for a $E_\nu^{-2}$ spectrum \citep{2020PhRvL.124e1103A}. 
The next-generation detector IceCube-Gen2 will have about five times better sensitivity than IceCube \citep{2020arXiv200804323T}. Although the point-source neutrino detection of SS 433 would still be challenging, it might be possible with more than two decades of the observations by IceCube-Gen2, or joint-analyses with HAWC-like detectors would be useful. 
If the proton spectrum is harder, the neutrino detection can be more promising. However, in such case, the gamma-ray model spectrum would become inconsistent with the {\it Fermi} LAT data and the existing upper limits by HESS/MAGIC. Using the HAWC data and the HESS/MAGIC upper limit, \citet{2019APh...109...25R} reached the similar conclusion, arguing that detectable neutrinos could be emitted from the inner region.

\subsection{Particle acceleration efficiency in other objects}

The SS 433 jets have a low $\eta$, while jets in radio galaxies may have various values of $\eta$. Other cosmic-ray accelerators generally have low values of $\eta$. Sharp X-ray images are observed from the forward shocks in supernova remnants (SNRs; \citealt{2005ApJ...621..793B}). The cutoff frequency in SNRs are $10^{17}$ Hz $-10^{18}$ Hz, and shock velocities are $\sim2000-10^4\rm~km~s^{-1}$ \citep{2008ARA&A..46...89R}. These values require $\eta\sim1$, according to Equation (\ref{eq:eta}). Also, fittings of pulsar wind nebulae (PWNe) demand a very efficient particle acceleration of $\eta\sim1$ \citep{2011ApJ...741...40T,2013MNRAS.429.2945T}.

There are a few possible reasons for such distinct values of $\eta$. To achieve a high value of $\eta$, strong turbulence should exist. A possible mechanism generating turbulence is density perturbations in the upstream of the shocks. ISM in our Galaxy has strong density perturbations, which can drive strong turbulence when shocks sweep up the ISM \citep{IYI12a,2019ApJ...886...54T}. The hotspots and the blazar emission regions can arise at the reverse shocks in expanding jets, which likely have weaker density perturbations due to adiabatic expansion. Another possibility is related to the plasma composition. Magnetized ion-electron plasmas result in strong turbulence owing to the streaming instability \citep{1975MNRAS.173..255S,Bel04a}, whereas electron-positron pair plasmas may not trigger it. However, PWNe likely accelerate particles at reverse shocks and consist of pair plasmas, and neither of the interpretations are applicable. Further studies are necessary on both the theoretical and observational sides to understand the dichotomy of the acceleration efficiency.

\subsection{Effects of escaping CRs}

In our hadronic scenarios, the protons of PeV energies escape from the system and are injected into the ISM. The diffusion coefficient in the ISM is often estimated by the Boron-to-Carbon ratio to be $D_{\rm ISM}\approx3\times10^{30}E_{p,\rm~PeV}^{1/3}\rm~cm^2~s^{-1}$, where $E_{p,\rm PeV}=E_p/(1\rm~PeV)$ and we consider the Kolmogorov turbulence \citep{SMP07a}. Then, the diffusion length during the lifetime of SS 433, $t_{\rm age}$, is estimated to be 
\begin{equation}
R_{\rm diff}\approx\sqrt{6D_{\rm ISM}t_{\rm age}}\simeq1.4E_{p,\rm PeV}^{1/6}t_{\rm age,12}^{1/2}\rm~ kpc.
\end{equation}
Since $R_{\rm diff}$ is shorter than the distance to the Earth, the CRs escaping from SS 433 have not arrived at Earth yet. The energy density of the escaping CRs at the PeV energy is estimated to be 
\begin{eqnarray}
& &U_{\rm CR,PeV}\approx\frac{3\epsilon_pL_jt_{\rm age}f_{\rm bol}}{4\pi R_{\rm diff}^3}\nonumber\\
&\sim&3\times10^{-5}\epsilon_{p,-1}L_{j,39.3}t_{\rm age,12}^{-1/2}E_{p,\rm PeV}^{-1} \rm~eV~cm^{-3},
\end{eqnarray}
where the factor $f_{\rm bol}\approx1/\ln(E_{p,\rm max}/\rm GeV)\sim1/15$ is the bolometric correction factor. The recent observations by Telescope Array Low-energy Extension (TALE) and IceTop reported that the CR energy density at the PeV energy is $\sim1\times10^{-4}\rm~eV~cm^{-3}$ \citep{TA18a,2019PhRvD.100h2002A}, which matches the estimate above within an order of magnitude. Hence, Galactic X-ray binaries may provide some contribution to the PeV CRs \citep[cf.,][]{2020MNRAS.493.3212C}.  The lifetime of SS 433 may be longer, $t_{\rm age}\sim10^5-10^6$ yr \citep{2008PASJ...60..715Y,2018ApJ...863..103S}, and the escaping CRs can arrive at Earth if we use $t_{\rm age}=10^6$ yr. In this case, SS 433 can contribute to the observed PeV CRs up to 6 \%.

\subsection{Comparison to previous work}
Previous studies on TeV gamma-ray emission from SS 433 mainly discussed the leptonic scenarios \citep{2018Natur.562...82A,2019ApJ...872...25X,2020ApJ...889..146S}. Our leptonic scenarios C and D are similar to the models by \cite{2020ApJ...889..146S} and \cite{2018Natur.562...82A}, respectively. 
However, we find that scenario C cannot reproduce the latest {\it Fermi} data by \cite{2020ApJ...889L...5F}, although the GeV detection is not significant enough by the {\it Fermi} data alone. On the other hand, scenario D can reproduce the {\it Fermi} data as argued in \cite{2020ApJ...889L...5F}.

The previous literature concluded that the hadronic scenarios are disfavored because the required jet power is too high for a typical number density of $\sim0.01-0.1\rm~cm^{-3}$ in the W50 nebula \citep{2018Natur.562...82A,2020ApJ...889..146S}. However, their conclusions were obtained without examining the effect of optical filaments, where the density can be much higher. A higher number density enables our hadronic scenario to naturally reproduce the GeV--TeV gamma-ray data with a reasonable jet power. Therefore, we conclude that both leptonic and hadronic scenarios can reproduce the GeV--TeV gamma-ray data.

\section{Summary}\label{sec:summary}

\begin{table}
\begin{center}
\caption{Consistency check for our scenarios. Here ``AGN analog'' indicates whether the value of $\eta$ in our scenario is consistent with those obtained from the radio galaxies' X-ray knots. $\bigcirc$, $\triangle$, and $\times$ indicate consistent, marginal, and inconsistent, respectively.}\label{tab:consistency}
\begin{tabular}{|c|cc|cc|}
\hline
 & Hadronic & & Leptonic & \\
Scenario & A   & B  & C  & D  \\
 & Knot & Hotspot & Knot & Hotspot\\
\hline
HAWC data & $\bigcirc$ & $\bigcirc$ & $\bigcirc$ & $\bigcirc$ \\
{\it Fermi} data & $\bigcirc$ & $\bigcirc$ & $\times$ & $\bigcirc$ \\
Ambient density & $\triangle$ & $\bigcirc$ & $\bigcirc$ & $\bigcirc$ \\
AGN analog & $\bigcirc$ & $\times$ & $\bigcirc$ & $\times$ \\
\hline
\end{tabular}
\end{center} 
\end{table}

We examined both leptonic and hadronic scenarios for GeV-TeV gamma-ray emission from the SS 433 jets in light of the recent detections by {\it Fermi} and HAWC. The gamma-ray emission region coincides with the X-ray knots and the optical filaments, where particle acceleration should be efficient and the target density should be high, respectively. To obtain broadband photon spectra, we solved the transport equations for electrons and protons taking into account acceleration, radiative and adiabatic cooling, and diffusive and advective escape. Fixing several parameters based on the multi-wavelength observations of the SS 433/W50 system, we searched parameter sets with which the resulting photon spectra match the observed data. We found that both hadronic and leptonic scenarios can reproduce the observed data without violating current observational constraints. The radio to X-ray data are emitted by electron synchrotron radiation and the GeV--TeV gamma-rays are produced by either the pion decay process or IC emission. The spectral shapes strongly depend on the advection timescale, and future observations by CTA, LHAASO and SWGO will provide more clues to distinguish between the scenarios.

Finally, we summarize the feasibility of our scenarios in terms of the gamma-ray spectrum, the ambient number density, and analogy to large-scale AGN jets (see Table \ref{tab:consistency}). Scenarios A, B, and D can reproduce the GeV-TeV gamma-ray data, while the scenario C cannot reproduce the {\it Fermi} data.  
The estimates of the ambient density in the W50 region prefer $n_{\rm eff}\sim0.01-0.1\rm~cm^{-3}$ \citep{1999ApJ...512..784S,2017A&A...599A..77P}, which is consistent with scenarios B, C, and D. 
However, the density in the optical filaments is as high as $100\rm~cm^{-3}$, and the filling factor of the filaments is unclear from observations. Thus, a value for scenario A of $n_{\rm eff}\sim10~\rm cm^{-3}$ is also acceptable.
In the large-scale jets of radio galaxies, the knots and hotspots have low and high values of $\eta$, respectively. Our scenarios assume a low value of $\eta$, which corresponds to the values in knots where the advection velocity is high, making scenarios A and C suitable.
Therefore, in this regard, we conclude that our hadronic scenario A would be the most plausible scenario for the high-energy gamma-ray emission mechanism of SS433. If the same mechanism operates in radio galaxies this implies that the X-ray knot region of the jets in radio galaxies may accelerate heavy nuclei up to ultrahigh energies. To more solidly understand the emission mechanisms in these objects, further investigations from both the observational and theoretical sides are necessary. In particular, future MeV gamma-ray observations will clarify the value of $\eta$ and observations of $>100$ TeV photons by LHAASSO, SWGO or CTA may be able to discriminate between the scenarios.


\begin{acknowledgements}
We thank  Ke Fang,  Takahiro Sudoh, and Kenji Toma for useful discussion. This work is partly supported by  JSPS Research Fellowship, KAKENHI Nos. 19J00198 (S.S.K.), and the Alfred P. Sloan Foundation, NSF Grant No.~AST-1908689, and KAKENHI No.~20H01901 (K.M.), and the Eberly Foundation (P.M.).
\end{acknowledgements}

\bibliographystyle{aasjournal}
\bibliography{ssk}

\begin{thebibliography}{}
\expandafter\ifx\csname natexlab\endcsname\relax\def\natexlab#1{#1}\fi
\providecommand{\url}[1]{\href{#1}{#1}}
\providecommand{\dodoi}[1]{doi:~\href{http://doi.org/#1}{\nolinkurl{#1}}}
\providecommand{\doeprint}[1]{\href{http://ascl.net/#1}{\nolinkurl{http://ascl.net/#1}}}
\providecommand{\doarXiv}[1]{\href{https://arxiv.org/abs/#1}{\nolinkurl{https://arxiv.org/abs/#1}}}

\bibitem[{Aab {et~al.}(2014)}]{Aab:2014kda}
Aab, A., {et~al.} 2014, Phys. Rev., D90, 122005,
  \dodoi{10.1103/PhysRevD.90.122005}

\bibitem[{{Aartsen} {et~al.}(2019){Aartsen}, {Ackermann}, {Adams}, {Aguilar},
  {Ahlers}, {Ahrens}, {Alispach}, {Andeen}, {Anderson}, {Ansseau}, {Anton},
  {Arg{\"u}elles}, {Auffenberg}, {Axani}, {Backes}, {Bagherpour}, {Bai},
  {Barbano}, {Barwick}, {Baum}, {Baur}, {Bay}, {Beatty}, {Becker}, {Becker
  Tjus}, {BenZvi}, {Berley}, {Bernardini}, {Besson}, {Binder}, {Bindig},
  {Blaufuss}, {Blot}, {Bohm}, {B{\"o}rner}, {B{\"o}ser}, {Botner},
  {B{\"o}ttcher}, {Bourbeau}, {Bourbeau}, {Bradascio}, {Braun}, {Bretz},
  {Bron}, {Brostean-Kaiser}, {Burgman}, {Buscher}, {Busse}, {Carver}, {Chen},
  {Cheung}, {Chirkin}, {Clark}, {Classen}, {Collin}, {Conrad}, {Coppin},
  {Correa}, {Cowen}, {Cross}, {Dave}, {de Andr{\'e}}, {De Clercq}, {DeLaunay},
  {Dembinski}, {Deoskar}, {De Ridder}, {Desiati}, {de Vries}, {de Wasseige},
  {de With}, {DeYoung}, {Diaz}, {D{\'\i}az-V{\'e}lez}, {Dujmovic}, {Dunkman},
  {Dvorak}, {Eberhardt}, {Ehrhardt}, {Eller}, {Evenson}, {Fahey}, {Fazely},
  {Felde}, {Feusels}, {Filimonov}, {Finley}, {Franckowiak}, {Friedman},
  {Fritz}, {Gaisser}, {Gallagher}, {Ganster}, {Garrappa}, {Gerhardt},
  {Ghorbani}, {Glauch}, {Gl{\"u}senkamp}, {Goldschmidt}, {Gonzalez}, {Grant},
  {Griffith}, {G{\"u}nder}, {G{\"u}nd{\"u}z}, {Haack}, {Hallgren}, {Halve},
  {Halzen}, {Hanson}, {Hebecker}, {Heereman}, {Heix}, {Helbing}, {Hellauer},
  {Henningsen}, {Hickford}, {Hignight}, {Hill}, {Hoffman}, {Hoffmann},
  {Hoinka}, {Hokanson-Fasig}, {Hoshina}, {Huang}, {Huber}, {Hultqvist},
  {H{\"u}nnefeld}, {Hussain}, {In}, {Iovine}, {Ishihara}, {Jacobi},
  {Japaridze}, {Jeong}, {Jero}, {Jones}, {Jonske}, {Joppe}, {Kang}, {Kappes},
  {Kappesser}, {Karg}, {Karl}, {Karle}, {Katz}, {Kauer}, {Kelley},
  {Kheirandish}, {Kim}, {Kintscher}, {Kiryluk}, {Kittler}, {Klein}, {Koirala},
  {Kolanoski}, {K{\"o}pke}, {Kopper}, {Kopper}, {Koskinen}, {Kowalski},
  {Krings}, {Kr{\"u}ckl}, {Kulacz}, {Kunwar}, {Kurahashi}, {Kyriacou},
  {Labare}, {Lanfranchi}, {Larson}, {Lauber}, {Lazar}, {Leonard}, {Leuermann},
  {Liu}, {Lohfink}, {Lozano Mariscal}, {Lu}, {Lucarelli}, {L{\"u}nemann},
  {Luszczak}, {Madsen}, {Maggi}, {Mahn}, {Makino}, {Mallik}, {Mallot},
  {Mancina}, {Mari{\textcommabelow s}}, {Maruyama}, {Mase}, {Maunu}, {Meagher},
  {Medici}, {Medina}, {Meier}, {Meighen-Berger}, {Menne}, {Merino}, {Meures},
  {Miarecki}, {Micallef}, {Moment{\'e}}, {Montaruli}, {Moore}, {Morse},
  {Moulai}, {Muth}, {Nagai}, {Nahnhauer}, {Nakarmi}, {Naumann}, {Neer},
  {Niederhausen}, {Nowicki}, {Nygren}, {Obertacke Pollmann}, {Olivas},
  {O'Murchadha}, {O'Sullivan}, {Palczewski}, {Pandya}, {Pankova}, {Park},
  {Peiffer}, {P{\'e}rez de los Heros}, {Philippen}, {Pieloth}, {Pinat},
  {Pizzuto}, {Plum}, {Porcelli}, {Price}, {Przybylski}, {Raab}, {Raissi},
  {Rameez}, {Rauch}, {Rawlins}, {Rea}, {Reimann}, {Relethford}, {Renzi},
  {Resconi}, {Rhode}, {Richman}, {Robertson}, {Rongen}, {Rott}, {Ruhe},
  {Ryckbosch}, {Rysewyk}, {Safa}, {Sanchez Herrera}, {Sandrock}, {Sand roos},
  {Santander}, {Sarkar}, {Sarkar}, {Satalecka}, {Schaufel}, {Schlunder},
  {Schmidt}, {Schneider}, {Schneider}, {Schumacher}, {Sclafani}, {Seckel},
  {Seunarine}, {Shefali}, {Silva}, {Snihur}, {Soedingrekso}, {Soldin}, {Song},
  {Spiczak}, {Spiering}, {Stachurska}, {Stamatikos}, {Stanev}, {Stasik},
  {Stein}, {Stettner}, {Steuer}, {Stezelberger}, {Stokstad}, {St{\"o}{\ss}l},
  {Strotjohann}, {St{\"u}rwald}, {Stuttard}, {Sullivan}, {Sutherland},
  {Taboada}, {Tenholt}, {Ter-Antonyan}, {Terliuk}, {Tilav}, {Tomankova},
  {T{\"o}nnis}, {Toscano}, {Tosi}, {Tselengidou}, {Tung}, {Turcati},
  {Turcotte}, {Turley}, {Ty}, {Unger}, {Unland Elorrieta}, {Usner},
  {Vandenbroucke}, {Van Driessche}, {van Eijk}, {van Eijndhoven}, {Vanheule},
  {van Santen}, {Vraeghe}, {Walck}, {Wallace}, {Wallraff}, {Wandkowsky},
  {Watson}, {Weaver}, {Weiss}, {Weldert}, {Wendt}, {Werthebach}, {Westerhoff},
  {Whelan}, {Whitehorn}, {Wiebe}, {Wiebusch}, {Wille}, {Williams}, {Wills},
  {Wolf}, {Wood}, {Wood}, {Woschnagg}, {Wrede}, {Xu}, {Xu}, {Xu}, {Yanez},
  {Yodh}, {Yoshida}, {Yuan}, {Z{\"o}cklein}, \& {IceCube
  Collaboration}}]{2019PhRvD.100h2002A}
{Aartsen}, M.~G., {Ackermann}, M., {Adams}, J., {et~al.} 2019, \prd, 100,
  082002, \dodoi{10.1103/PhysRevD.100.082002}

\bibitem[{{Aartsen} {et~al.}(2020){Aartsen}, {Ackermann}, {Adams}, {Aguilar},
  {Ahlers}, {Ahrens}, {Alispach}, {Andeen}, {Anderson}, {Ansseau}, {Anton},
  {Arg{\"u}elles}, {Auffenberg}, {Axani}, {Backes}, {Bagherpour}, {Bai},
  {Balagopal}, {Barbano}, {Barwick}, {Bastian}, {Baum}, {Baur}, {Bay},
  {Beatty}, {Becker}, {Becker Tjus}, {BenZvi}, {Berley}, {Bernardini},
  {Besson}, {Binder}, {Bindig}, {Blaufuss}, {Blot}, {Bohm}, {B{\"o}rner},
  {B{\"o}ser}, {Botner}, {B{\"o}ttcher}, {Bourbeau}, {Bourbeau}, {Bradascio},
  {Braun}, {Bron}, {Brostean-Kaiser}, {Burgman}, {Buscher}, {Busse}, {Carver},
  {Chen}, {Cheung}, {Chirkin}, {Choi}, {Clark}, {Classen}, {Coleman}, {Collin},
  {Conrad}, {Coppin}, {Correa}, {Cowen}, {Cross}, {Dave}, {De Clercq},
  {DeLaunay}, {Dembinski}, {Deoskar}, {De Ridder}, {Desiati}, {de Vries}, {de
  Wasseige}, {de With}, {DeYoung}, {Diaz}, {D{\'\i}az-V{\'e}lez}, {Dujmovic},
  {Dunkman}, {Dvorak}, {Eberhardt}, {Ehrhardt}, {Eller}, {Engel}, {Evenson},
  {Fahey}, {Fazely}, {Felde}, {Filimonov}, {Finley}, {Fox}, {Franckowiak},
  {Friedman}, {Fritz}, {Gaisser}, {Gallagher}, {Ganster}, {Garrappa},
  {Gerhardt}, {Ghorbani}, {Glauch}, {Gl{\"u}senkamp}, {Goldschmidt},
  {Gonzalez}, {Grant}, {Griffith}, {Griswold}, {G{\"u}nder}, {G{\"u}nd{\"u}z},
  {Haack}, {Hallgren}, {Halliday}, {Halve}, {Halzen}, {Hanson}, {Haungs},
  {Hebecker}, {Heereman}, {Heix}, {Helbing}, {Hellauer}, {Henningsen},
  {Hickford}, {Hignight}, {Hill}, {Hoffman}, {Hoffmann}, {Hoinka},
  {Hokanson-Fasig}, {Hoshina}, {Huang}, {Huber}, {Huber}, {Hultqvist},
  {H{\"u}nnefeld}, {Hussain}, {In}, {Iovine}, {Ishihara}, {Japaridze}, {Jeong},
  {Jero}, {Jones}, {Jonske}, {Joppe}, {Kang}, {Kang}, {Kappes}, {Kappesser},
  {Karg}, {Karl}, {Karle}, {Katz}, {Kauer}, {Kelley}, {Kheirandish}, {Kim},
  {Kintscher}, {Kiryluk}, {Kittler}, {Klein}, {Koirala}, {Kolanoski},
  {K{\"o}pke}, {Kopper}, {Kopper}, {Koskinen}, {Kowalski}, {Krings},
  {Kr{\"u}ckl}, {Kulacz}, {Kurahashi}, {Kyriacou}, {Labare}, {Lanfranchi},
  {Larson}, {Lauber}, {Lazar}, {Leonard}, {Leszczy{\'n}ska}, {Leuermann},
  {Liu}, {Lohfink}, {Lozano Mariscal}, {Lu}, {Lucarelli}, {L{\"u}nemann},
  {Luszczak}, {Lyu}, {Ma}, {Madsen}, {Maggi}, {Mahn}, {Makino}, {Mallik},
  {Mallot}, {Mancina}, {Mari{\textcommabelow s}}, {Maruyama}, {Mase}, {Matis},
  {Maunu}, {McNally}, {Meagher}, {Medici}, {Medina}, {Meier}, {Meighen-Berger},
  {Menne}, {Merino}, {Meures}, {Micallef}, {Mockler}, {Moment{\'e}},
  {Montaruli}, {Moore}, {Morse}, {Moulai}, {Muth}, {Nagai}, {Naumann}, {Neer},
  {Niederhausen}, {Nisa}, {Nowicki}, {Nygren}, {Obertacke Pollmann}, {Oehler},
  {Olivas}, {O'Murchadha}, {O'Sullivan}, {Palczewski}, {Pandya}, {Pankova},
  {Park}, {Peiffer}, {P{\'e}rez de los Heros}, {Philippen}, {Pieloth}, {Pinat},
  {Pizzuto}, {Plum}, {Porcelli}, {Price}, {Przybylski}, {Raab}, {Raissi},
  {Rameez}, {Rauch}, {Rawlins}, {Rea}, {Reimann}, {Relethford}, {Renschler},
  {Renzi}, {Resconi}, {Rhode}, {Richman}, {Robertson}, {Rongen}, {Rott},
  {Ruhe}, {Ryckbosch}, {Rysewyk}, {Safa}, {Sanchez Herrera}, {Sandrock},
  {Sandroos}, {Santander}, {Sarkar}, {Sarkar}, {Satalecka}, {Schaufel},
  {Schieler}, {Schlunder}, {Schmidt}, {Schneider}, {Schneider}, {Schr{\"o}der},
  {Schumacher}, {Sclafani}, {Seckel}, {Seunarine}, {Shefali}, {Silva},
  {Snihur}, {Soedingrekso}, {Soldin}, {Song}, {Spiczak}, {Spiering},
  {Stachurska}, {Stamatikos}, {Stanev}, {Stein}, {Steinm{\"u}ller}, {Stettner},
  {Steuer}, {Stezelberger}, {Stokstad}, {St{\"o}{\ss}l}, {Strotjohann},
  {St{\"u}rwald}, {Stuttard}, {Sullivan}, {Taboada}, {Tenholt}, {Ter-Antonyan},
  {Terliuk}, {Tilav}, {Tollefson}, {Tomankova}, {T{\"o}nnis}, {Toscano},
  {Tosi}, {Trettin}, {Tselengidou}, {Tung}, {Turcati}, {Turcotte}, {Turley},
  {Ty}, {Unger}, {Unland Elorrieta}, {Usner}, {Vandenbroucke}, {Van Driessche},
  {van Eijk}, {van Eijndhoven}, {Vanheule}, {van Santen}, {Vraeghe}, {Walck},
  {Wallace}, {Wallraff}, {Wandkowsky}, {Watson}, {Weaver}, {Weindl}, {Weiss},
  {Weldert}, {Wendt}, {Werthebach}, {Whelan}, {Whitehorn}, {Wiebe}, {Wiebusch},
  {Wille}, {Williams}, {Wills}, {Wolf}, {Wood}, {Wood}, {Woschnagg}, {Wrede},
  {Xu}, {Xu}, {Xu}, {Yanez}, {Yodh}, {Yoshida}, {Yuan}, \&
  {Z{\"o}cklein}}]{2020PhRvL.124e1103A}
---. 2020, \prl, 124, 051103, \dodoi{10.1103/PhysRevLett.124.051103}

\bibitem[{{Abbasi} {et~al.}(2018){Abbasi}, {Abe}, {Abu-Zayyad}, {Allen},
  {Azuma}, {Barcikowski}, {Belz}, {Bergman}, {Blake}, {Cady}, {Cheon}, {Chiba},
  {Chikawa}, {Di Matteo}, {Fujii}, {Fujita}, {Fukushima}, {Furlich}, {Goto},
  {Hanlon}, {Hayashi}, {Hayashi}, {Hayashida}, {Hibino}, {Honda}, {Ikeda},
  {Inoue}, {Ishii}, {Ishimori}, {Ito}, {Ivanov}, {Jeong}, {Jeong}, {Jui},
  {Kadota}, {Kakimoto}, {Kalashev}, {Kasahara}, {Kawai}, {Kawakami}, {Kawana},
  {Kawata}, {Kido}, {Kim}, {Kim}, {Kim}, {Kishigami}, {Kitamura}, {Kitamura},
  {Kuzmin}, {Kuznetsov}, {Kwon}, {Lee}, {Lubsandorzhiev}, {Lundquist},
  {Machida}, {Martens}, {Matsuyama}, {Matthews}, {Mayta}, {Minamino}, {Mukai},
  {Myers}, {Nagasawa}, {Nagataki}, {Nakamura}, {Nakamura}, {Nonaka}, {Nozato},
  {Oda}, {Ogio}, {Ogura}, {Ohnishi}, {Ohoka}, {Okuda}, {Omura}, {Ono}, {Onogi},
  {Oshima}, {Ozawa}, {Park}, {Pshirkov}, {Rodriguez}, {Rubtsov}, {Ryu},
  {Sagawa}, {Sahara}, {Saito}, {Saito}, {Sakaki}, {Sakurai}, {Scott}, {Seki},
  {Sekino}, {Shah}, {Shibata}, {Shibata}, {Shimodaira}, {Shin}, {Shin},
  {Smith}, {Sokolsky}, {Stokes}, {Stratton}, {Stroman}, {Suzawa}, {Takagi},
  {Takahashi}, {Takamura}, {Takeda}, {Takeishi}, {Taketa}, {Takita}, {Tameda},
  {Tanaka}, {Tanaka}, {Tanaka}, {Thomas}, {Thomson}, {Tinyakov}, {Tkachev},
  {Tokuno}, {Tomida}, {Troitsky}, {Tsunesada}, {Tsutsumi}, {Uchihori}, {Udo},
  {Urban}, {Wong}, {Yamamoto}, {Yamane}, {Yamaoka}, {Yamazaki}, {Yang},
  {Yashiro}, {Yoneda}, {Yoshida}, {Yoshii}, {Zhezher}, {Zundel}, \& {Telescope
  Array Collaboration}}]{TA18a}
{Abbasi}, R.~U., {Abe}, M., {Abu-Zayyad}, T., {et~al.} 2018, \apj, 865, 74,
  \dodoi{10.3847/1538-4357/aada05}

\bibitem[{{Abell} \& {Margon}(1979)}]{1979Natur.279..701A}
{Abell}, G.~O., \& {Margon}, B. 1979, \nat, 279, 701, \dodoi{10.1038/279701a0}

\bibitem[{{Abeysekara} {et~al.}(2018){Abeysekara}, {Albert}, {Alfaro},
  {Alvarez}, {{\'A}lvarez}, {Arceo}, {Arteaga-Vel{\'a}zquez}, {Avila Rojas},
  {Ayala Solares}, {Belmont-Moreno}, {BenZvi}, {Brisbois}, {Caballero-Mora},
  {Capistr{\'a}n}, {Carrami{\~n}ana}, {Casanova}, {Castillo}, {Cotti},
  {Cotzomi}, {Couti{\~n}o de Le{\'o}n}, {De Le{\'o}n}, {De la Fuente},
  {D{\'\i}az-V{\'e}lez}, {Dichiara}, {Dingus}, {DuVernois}, {Ellsworth},
  {Engel}, {Espinoza}, {Fang}, {Fleischhack}, {Fraija}, {Galv{\'a}n-G{\'a}mez},
  {Garc{\'\i}a-Gonz{\'a}lez}, {Garfias}, {Gonz{\'a}lez-Mu{\~n}oz},
  {Gonz{\'a}lez}, {Goodman}, {Hampel-Arias}, {Harding}, {Hernandez}, {Hinton},
  {Hona}, {Hueyotl-Zahuantitla}, {Hui}, {H{\"u}ntemeyer}, {Iriarte},
  {Jardin-Blicq}, {Joshi}, {Kaufmann}, {Kar}, {Kunde}, {Lauer}, {Lee},
  {Le{\'o}n Vargas}, {Li}, {Linnemann}, {Longinotti}, {Luis-Raya},
  {L{\'o}pez-Coto}, {Malone}, {Marinelli}, {Martinez}, {Martinez-Castellanos},
  {Mart{\'\i}nez-Castro}, {Matthews}, {Mirand a-Romagnoli}, {Moreno},
  {Mostaf{\'a}}, {Nayerhoda}, {Nellen}, {Newbold}, {Nisa}, {Noriega-Papaqui},
  {Pretz}, {P{\'e}rez-P{\'e}rez}, {Ren}, {Rho}, {Rivi{\`e}re},
  {Rosa-Gonz{\'a}lez}, {Rosenberg}, {Ruiz-Velasco}, {Salesa Greus}, {Sandoval},
  {Schneider}, {Schoorlemmer}, {Seglar Arroyo}, {Sinnis}, {Smith}, {Springer},
  {Surajbali}, {Taboada}, {Tibolla}, {Tollefson}, {Torres}, {Vianello},
  {Villase{\~n}or}, {Weisgarber}, {Werner}, {Westerhoff}, {Wood}, {Yapici},
  {Yodh}, {Zepeda}, {Zhang}, \& {Zhou}}]{2018Natur.562...82A}
{Abeysekara}, A.~U., {Albert}, A., {Alfaro}, R., {et~al.} 2018, \nat, 562, 82,
  \dodoi{10.1038/s41586-018-0565-5}

\bibitem[{{Albert} {et~al.}(2019){Albert}, {Alfaro}, {Ashkar}, {Alvarez},
  {{\'A}lvarez}, {Arteaga-Vel{\'a}zquez}, {Ayala Solares}, {Arceo}, {Bellido},
  {BenZvi}, {Bretz}, {Brisbois}, {Brown}, {Brun}, {Caballero-Mora}, {Carosi},
  {Carrami{\~n}ana}, {Casanova}, {Chadwick}, {Cotter}, {Couti{\~n}o De
  Le{\'o}n}, {Cristofari}, {Dasso}, {de la Fuente}, {Dingus}, {Desiati},
  {Salles}, {de Souza}, {Dorner}, {D{\'\i}az-V{\'e}lez},
  {Garc{\'\i}a-Gonz{\'a}lez}, {DuVernois}, {Di Sciascio}, {Engel},
  {Fleischhack}, {Fraija}, {Funk}, {Glicenstein}, {Gonzalez}, {Gonz{\'a}lez},
  {Goodman}, {Harding}, {Haungs}, {Hinton}, {Hona}, {Hoyos}, {Huentemeyer},
  {Iriarte}, {Jardin-Blicq}, {Joshi}, {Kaufmann}, {Kawata}, {Kunwar},
  {Lefaucheur}, {Lenain}, {Link}, {L{\'o}pez-Coto}, {Marandon}, {Mariotti},
  {Mart{\'\i}nez-Castro}, {Mart{\'\i}nez-Huerta}, {Mostaf{\'a}}, {Nayerhoda},
  {Nellen}, {de O{\~n}a Wilhelmi}, {Parsons}, {Patricelli}, {Pichel}, {Piel},
  {Prandini}, {Pueschel}, {Procureur}, {Reisenegger}, {Rivi{\`e}re},
  {Rodriguez}, {Rovero}, {Rowell}, {Ruiz-Velasco}, {Sandoval}, {Santander},
  {Sako}, {Sako}, {Satalecka}, {Schoorlemmer}, {Sch{\"u}ssler},
  {Seglar-Arroyo}, {Smith}, {Spencer}, {Surajbali}, {Tabachnick}, {Taylor},
  {Tibolla}, {Torres}, {Vallage}, {Viana}, {Watson}, {Weisgarber}, {Werner},
  {White}, {Wischnewski}, {Yang}, {Zepeda}, \& {Zhou}}]{2019arXiv190208429A}
{Albert}, A., {Alfaro}, R., {Ashkar}, H., {et~al.} 2019, arXiv e-prints,
  arXiv:1902.08429.
\newblock \doarXiv{1902.08429}

\bibitem[{{Aramaki} {et~al.}(2020){Aramaki}, {Adrian}, {Karagiorgi}, \&
  {Odaka}}]{2020APh...114..107A}
{Aramaki}, T., {Adrian}, P. O.~H., {Karagiorgi}, G., \& {Odaka}, H. 2020,
  Astroparticle Physics, 114, 107, \dodoi{10.1016/j.astropartphys.2019.07.002}

\bibitem[{{Araudo} {et~al.}(2016){Araudo}, {Bell}, {Crilly}, \&
  {Blundell}}]{2016MNRAS.460.3554A}
{Araudo}, A.~T., {Bell}, A.~R., {Crilly}, A., \& {Blundell}, K.~M. 2016,
  \mnras, 460, 3554, \dodoi{10.1093/mnras/stw1204}

\bibitem[{{Arshakian} \& {Longair}(2004)}]{2004MNRAS.351..727A}
{Arshakian}, T.~G., \& {Longair}, M.~S. 2004, \mnras, 351, 727,
  \dodoi{10.1111/j.1365-2966.2004.07823.x}

\bibitem[{{Atoyan} \& {Dermer}(2004)}]{2004ApJ...613..151A}
{Atoyan}, A., \& {Dermer}, C.~D. 2004, \apj, 613, 151, \dodoi{10.1086/422499}

\bibitem[{{Bai} {et~al.}(2019){Bai}, {Bi}, {Bi}, {Cao}, {Chen}, {Chen},
  {Chiavassa}, {Cui}, {Dai}, {della Volpe}, {Di Girolamo}, {Di Sciascio},
  {Fan}, {Giacalone}, {Guo}, {He}, {He}, {Heller}, {Huang}, {Huang}, {Jia},
  {Ksenofontov}, {Leahy}, {Li}, {Li}, {Liang}, {Lipari}, {Liu}, {Liu}, {Liu},
  {Ma}, {Martineau-Huynh}, {Martraire}, {Montaruli}, {Ruffolo}, {Stenkin},
  {Su}, {Tam}, {Tang}, {Tian}, {Vallania}, {Vernetto}, {Vigorito}, {Wang},
  {Wang}, {Wang}, {Wang}, {Wang}, {Wang}, {Wei}, {Wei}, {Wu}, {Wu}, {Wu},
  {Yan}, {Yang}, {Yang}, {Yao}, {Yin}, {Yuan}, {Zhang}, {Zhang}, {Zhang},
  {Zhang}, {Zhang}, {Zhang}, {Zhao}, {Zhou}, {Zhu}, \&
  {Zhu}}]{2019arXiv190502773B}
{Bai}, X., {Bi}, B.~Y., {Bi}, X.~J., {et~al.} 2019, arXiv e-prints,
  arXiv:1905.02773.
\newblock \doarXiv{1905.02773}

\bibitem[{{Bamba} {et~al.}(2005){Bamba}, {Yamazaki}, {Yoshida}, {Terasawa}, \&
  {Koyama}}]{2005ApJ...621..793B}
{Bamba}, A., {Yamazaki}, R., {Yoshida}, T., {Terasawa}, T., \& {Koyama}, K.
  2005, \apj, 621, 793, \dodoi{10.1086/427620}

\bibitem[{{Baring} {et~al.}(2017){Baring}, {B{\"o}ttcher}, \&
  {Summerlin}}]{2017MNRAS.464.4875B}
{Baring}, M.~G., {B{\"o}ttcher}, M., \& {Summerlin}, E.~J. 2017, \mnras, 464,
  4875, \dodoi{10.1093/mnras/stw2344}

\bibitem[{{Becker} {et~al.}(2006){Becker}, {Le}, \& {Dermer}}]{bld06}
{Becker}, P.~A., {Le}, T., \& {Dermer}, C.~D. 2006, \apj, 647, 539,
  \dodoi{10.1086/505319}

\bibitem[{{Bell}(1978)}]{Bel78a}
{Bell}, A.~R. 1978, \mnras, 182, 147, \dodoi{10.1093/mnras/182.2.147}

\bibitem[{{Bell}(2004)}]{Bel04a}
---. 2004, \mnras, 353, 550, \dodoi{10.1111/j.1365-2966.2004.08097.x}

\bibitem[{{Blandford} \& {Ostriker}(1978)}]{BO78a}
{Blandford}, R.~D., \& {Ostriker}, J.~P. 1978, \apjl, 221, L29,
  \dodoi{10.1086/182658}

\bibitem[{{Blumenthal} \& {Gould}(1970)}]{1970RvMP...42..237B}
{Blumenthal}, G.~R., \& {Gould}, R.~J. 1970, Reviews of Modern Physics, 42,
  237, \dodoi{10.1103/RevModPhys.42.237}

\bibitem[{{Bordas} {et~al.}(2015){Bordas}, {Yang}, {Kafexhiu}, \&
  {Aharonian}}]{2015ApJ...807L...8B}
{Bordas}, P., {Yang}, R., {Kafexhiu}, E., \& {Aharonian}, F. 2015, \apjl, 807,
  L8, \dodoi{10.1088/2041-8205/807/1/L8}

\bibitem[{{Boumis} {et~al.}(2007){Boumis}, {Meaburn}, {Alikakos}, {Redman},
  {Akras}, {Mavromatakis}, {L{\'o}pez}, {Caulet}, \&
  {Goudis}}]{2007MNRAS.381..308B}
{Boumis}, P., {Meaburn}, J., {Alikakos}, J., {et~al.} 2007, \mnras, 381, 308,
  \dodoi{10.1111/j.1365-2966.2007.12276.x}

\bibitem[{{Breiding} {et~al.}(2017){Breiding}, {Meyer}, {Georganopoulos},
  {Keenan}, {DeNigris}, \& {Hewitt}}]{2017ApJ...849...95B}
{Breiding}, P., {Meyer}, E.~T., {Georganopoulos}, M., {et~al.} 2017, \apj, 849,
  95, \dodoi{10.3847/1538-4357/aa907a}

\bibitem[{{Brinkmann} {et~al.}(2007){Brinkmann}, {Pratt}, {Rohr}, {Kawai}, \&
  {Burwitz}}]{2007A&A...463..611B}
{Brinkmann}, W., {Pratt}, G.~W., {Rohr}, S., {Kawai}, N., \& {Burwitz}, V.
  2007, \aap, 463, 611, \dodoi{10.1051/0004-6361:20065570}

\bibitem[{{Caprioli}(2015)}]{Cap15a}
{Caprioli}, D. 2015, \apjl, 811, L38, \dodoi{10.1088/2041-8205/811/2/L38}

\bibitem[{{Caprioli} \& {Spitkovsky}(2014)}]{CS14b}
{Caprioli}, D., \& {Spitkovsky}, A. 2014, \apj, 794, 46,
  \dodoi{10.1088/0004-637X/794/1/46}

\bibitem[{{Casse} \& {Marcowith}(2005)}]{2005APh....23...31C}
{Casse}, F., \& {Marcowith}, A. 2005, Astroparticle Physics, 23, 31,
  \dodoi{10.1016/j.astropartphys.2004.11.003}

\bibitem[{{Cherenkov Telescope Array Consortium} {et~al.}(2019){Cherenkov
  Telescope Array Consortium}, {Acharya}, {Agudo}, {Al Samarai}, {Alfaro},
  {Alfaro}, {Alispach}, {Alves Batista}, {Amans}, {Amato}, {Ambrosi},
  {Antolini}, {Antonelli}, {Aramo}, {Araya}, {Armstrong}, {Arqueros},
  {Arrabito}, {Asano}, {Ashley}, {Backes}, {Balazs}, {Balbo}, {Ballester},
  {Ballet}, {Bamba}, {Barkov}, {Barres de Almeida}, {Barrio}, {Bastieri},
  {Becherini}, {Belfiore}, {Benbow}, {Berge}, {Bernardini}, {Bernardini},
  {Bernardos}, {Bernl{\"o}hr}, {Bertucci}, {Biasuzzi}, {Bigongiari}, {Biland},
  {Bissaldi}, {Biteau}, {Blanch}, {Blazek}, {Boisson}, {Bolmont}, {Bonanno},
  {Bonardi}, {Bonavolont{\`a}}, {Bonnoli}, {Bosnjak}, {B{\"o}ttcher},
  {Braiding}, {Bregeon}, {Brill}, {Brown}, {Brun}, {Brunetti}, {Buanes},
  {Buckley}, {Bugaev}, {B{\"u}hler}, {Bulgarelli}, {Bulik}, {Burton},
  {Burtovoi}, {Busetto}, {Canestrari}, {Capalbi}, {Capitanio}, {Caproni},
  {Caraveo}, {C{\'a}rdenas}, {Carlile}, {Carosi}, {Carqu{\'\i}n}, {Carr},
  {Casanova}, {Cascone}, {Catalani}, {Catalano}, {Cauz}, {Cerruti}, {Chadwick},
  {Chaty}, {Chaves}, {Chen}, {Chen}, {Chernyakova}, {Chikawa}, {Christov},
  {Chudoba}, {Cie{\'s}lar}, {Coco}, {Colafrancesco}, {Colin}, {Conforti},
  {Connaughton}, {Conrad}, {Contreras}, {Cortina}, {Costa}, {Costantini},
  {Cotter}, {Covino}, {Crocker}, {Cuadra}, {Cuevas}, {Cumani}, {D'A{\`\i}},
  {D'Ammando}, {D'Avanzo}, {D'Urso}, {Daniel}, {Davids}, {Dawson}, {Dazzi}, {De
  Angelis}, {de C{\'a}ssia dos Anjos}, {De Cesare}, {De Franco}, {de Gouveia
  Dal Pino}, {de la Calle}, {de los Reyes Lopez}, {De Lotto}, {De Luca}, {De
  Lucia}, {de Naurois}, {de O{\~n}a Wilhelmi}, {De Palma}, {De Persio}, {de
  Souza}, {Deil}, {Del Santo}, {Delgado}, {della Volpe}, {Di Girolamo}, {Di
  Pierro}, {Di Venere}, {D{\'\i}az}, {Dib}, {Diebold}, {Djannati-Ata{\"\i}},
  {Dom{\'\i}nguez}, {Dominis Prester}, {Dorner}, {Doro}, {Drass}, {Dravins},
  {Dubus}, {Dwarkadas}, {Ebr}, {Eckner}, {Egberts}, {Einecke}, {Ekoume},
  {Els{\"a}sser}, {Ernenwein}, {Espinoza}, {Evoli}, {Fairbairn},
  {Falceta-Goncalves}, {Falcone}, {Farnier}, {Fasola}, {Fedorova}, {Fegan},
  {Fernand ez-Alonso}, {Fern{\'a}ndez-Barral}, {Ferrand}, {Fesquet},
  {Filipovic}, {Fioretti}, {Fontaine}, {Fornasa}, {Fortson}, {Freixas
  Coromina}, {Fruck}, {Fujita}, {Fukazawa}, {Funk}, {F{\"u}{\ss}ling},
  {Gabici}, {Gadola}, {Gallant}, {Garcia}, {Garcia L{\'o}pez}, {Garczarczyk},
  {Gaskins}, {Gasparetto}, {Gaug}, {Gerard}, {Giavitto}, {Giglietto}, {Giommi},
  {Giordano}, {Giro}, {Giroletti}, {Giuliani}, {Glicenstein}, {Gnatyk},
  {Godinovic}, {Goldoni}, {G{\'o}mez-Vargas}, {Gonz{\'a}lez}, {Gonz{\'a}lez},
  {G{\"o}tz}, {Graham}, {Grand i}, {Granot}, {Green}, {Greenshaw}, {Griffiths},
  {Gunji}, {Hadasch}, {Hara}, {Hardcastle}, {Hassan}, {Hayashi}, {Hayashida},
  {Heller}, {Helo}, {Hermann}, {Hinton}, {Hnatyk}, {Hofmann}, {Holder},
  {Horan}, {H{\"o}randel}, {Horns}, {Horvath}, {Hovatta}, {Hrabovsky},
  {Hrupec}, {Humensky}, {H{\"u}tten}, {Iarlori}, {Inada}, {Inome}, {Inoue},
  {Inoue}, {Inoue}, {Iocco}, {Ioka}, {Iori}, {Ishio}, {Iwamura}, {Jamrozy},
  {Janecek}, {Jankowsky}, {Jean}, {Jung-Richardt}, {Jurysek}, {Kaaret},
  {Karkar}, {Katagiri}, {Katz}, {Kawanaka}, {Kazanas}, {Kh{\'e}lifi}, {Kieda},
  {Kimeswenger}, {Kimura}, {Kisaka}, {Knapp}, {Kn{\"o}dlseder}, {Koch},
  {Kohri}, {Komin}, {Kosack}, {Kraus}, {Krause}, {Krau{\ss}}, {Kubo}, {Kukec
  Mezek}, {Kuroda}, {Kushida}, {La Palombara}, {Lamanna}, {Lang}, {Lapington},
  {Le Blanc}, {Leach}, {Lees}, {Lefaucheur}, {Leigui de Oliveira}, {Lenain},
  {Lico}, {Limon}, {Lindfors}, {Lohse}, {Lombardi}, {Longo}, {L{\'o}pez},
  {L{\'o}pez-Coto}, {Lu}, {Lucarelli}, {Luque-Escamilla}, {Lyard}, {Maccarone},
  {Maier}, {Majumdar}, {Malaguti}, {Mandat}, {Maneva}, {Manganaro}, {Mangano},
  {Marcowith}, {Mar{\'\i}n}, {Markoff}, {Mart{\'\i}}, {Martin},
  {Mart{\'\i}nez}, {Mart{\'\i}nez}, {Masetti}, {Masuda}, {Maurin}, {Maxted},
  {Mazin}, {Medina}, {Melandri}, {Mereghetti}, {Meyer}, {Minaya}, {Mirabal},
  {Mirzoyan}, {Mitchell}, {Mizuno}, {Moderski}, {Mohammed}, {Mohrmann},
  {Montaruli}, {Moralejo}, {Morcuende-Parrilla}, {Mori}, {Morlino}, {Morris},
  {Morselli}, {Moulin}, {Mukherjee}, {Mundell}, {Murach}, {Muraishi}, {Murase},
  {Nagai}, {Nagataki}, {Nagayoshi}, {Naito}, {Nakamori}, {Nakamura}, {Niemiec},
  {Nieto}, {Niko{\l}ajuk}, {Nishijima}, {Noda}, {Nosek}, {Novosyadlyj},
  {Nozaki}, {O'Brien}, {Oakes}, {Ohira}, {Ohishi}, {Ohm}, {Okazaki}, {Okumura},
  {Ong}, {Orienti}, {Orito}, {Osborne}, {Ostrowski}, {Otte}, {Oya}, {Padovani},
  {Paizis}, {Palatiello}, {Palatka}, {Paoletti}, {Paredes}, {Pareschi},
  {Parsons}, {Pe'er}, {Pech}, {Pedaletti}, {Perri}, {Persic}, {Petrashyk},
  {Petrucci}, {Petruk}, {Peyaud}, {Pfeifer}, {Piano}, {Pisarski}, {Pita},
  {Pohl}, {Polo}, {Pozo}, {Prandini}, {Prast}, {Principe}, {Prokhorov},
  {Prokoph}, {Prouza}, {P{\"u}hlhofer}, {Punch}, {P{\"u}rckhauer}, {Queiroz},
  {Quirrenbach}, {Rain{\`o}}, {Razzaque}, {Reimer}, {Reimer}, {Reisenegger},
  {Renaud}, {Rezaeian}, {Rhode}, {Ribeiro}, {Rib{\'o}}, {Richtler}, {Rico},
  {Rieger}, {Riquelme}, {Rivoire}, {Rizi}, {Rodriguez}, {Rodriguez Fernandez},
  {Rodr{\'\i}guez V{\'a}zquez}, {Rojas}, {Romano}, {Romeo}, {Rosado}, {Rovero},
  {Rowell}, {Rudak}, {Rugliancich}, {Rulten}, {Sadeh}, {Safi-Harb}, {Saito},
  {Sakaki}, {Sakurai}, {Salina}, {S{\'a}nchez-Conde}, {Sandaker}, {Sandoval},
  {Sangiorgi}, {Sanguillon}, {Sano}, {Santand er}, {Sarkar}, {Satalecka},
  {Saturni}, {Schioppa}, {Schlenstedt}, {Schneider}, {Schoorlemmer},
  {Schovanek}, {Schulz}, {Schussler}, {Schwanke}, {Sciacca}, {Scuderi},
  {Seitenzahl}, {Semikoz}, {Sergijenko}, {Servillat}, {Shalchi}, {Shellard},
  {Sidoli}, {Siejkowski}, {Sillanp{\"a}{\"a}}, {Sironi}, {Sitarek}, {Sliusar},
  {Slowikowska}, {Sol}, {Stamerra}, {Stani{\v{c}}}, {Starling}, {Stawarz},
  {Stefanik}, {Stephan}, {Stolarczyk}, {Stratta}, {Straumann}, {Suomijarvi},
  {Supanitsky}, {Tagliaferri}, {Tajima}, {Tavani}, {Tavecchio}, {Tavernet},
  {Tayabaly}, {Tejedor}, {Temnikov}, {Terada}, {Terrier}, {Terzic}, {Teshima},
  {Testa}, {Thoudam}, {Tian}, {Tibaldo}, {Tluczykont}, {Todero Peixoto},
  {Tokanai}, {Tomastik}, {Tonev}, {Tornikoski}, {Torres}, {Torresi}, {Tosti},
  {Tothill}, {Tovmassian}, {Travnicek}, {Trichard}, {Trifoglio}, {Troyano
  Pujadas}, {Tsujimoto}, {Umana}, {Vagelli}, {Vagnetti}, {Valentino},
  {Vallania}, {Valore}, {van Eldik}, {Vand enbroucke}, {Varner}, {Vasileiadis},
  {Vassiliev}, {V{\'a}zquez Acosta}, {Vecchi}, {Vega}, {Vercellone}, {Veres},
  {Vergani}, {Verzi}, {Vettolani}, {Viana}, {Vigorito}, {Villanueva}, {Voelk},
  {Vollhardt}, {Vorobiov}, {Vrastil}, {Vuillaume}, {Wagner}, {Wagner},
  {Walter}, {Ward}, {Warren}, {Watson}, {Werner}, {White}, {White},
  {Wierzcholska}, {Wilcox}, {Will}, {Williams}, {Wischnewski}, {Wood},
  {Yamamoto}, {Yamazaki}, {Yanagita}, {Yang}, {Yoshida}, {Yoshiike},
  {Yoshikoshi}, {Zacharias}, {Zaharijas}, {Zampieri}, {Zand anel}, {Zanin},
  {Zavrtanik}, {Zavrtanik}, {Zdziarski}, {Zech}, {Zechlin}, {Zhdanov},
  {Ziegler}, \& {Zorn}}]{2019scta.book.....C}
{Cherenkov Telescope Array Consortium}, {Acharya}, B.~S., {Agudo}, I., {et~al.}
  2019, {Science with the Cherenkov Telescope Array}, \dodoi{10.1142/10986}

\bibitem[{{Cooper} {et~al.}(2020){Cooper}, {Gaggero}, {Markoff}, \&
  {Zhang}}]{2020MNRAS.493.3212C}
{Cooper}, A.~J., {Gaggero}, D., {Markoff}, S., \& {Zhang}, S. 2020, \mnras,
  493, 3212, \dodoi{10.1093/mnras/staa373}

\bibitem[{{De Angelis} {et~al.}(2017){De Angelis}, {Tatischeff}, {Tavani},
  {Oberlack}, {Grenier}, {Hanlon}, {Walter}, {Argan}, {von Ballmoos}, \&
  {Bulgarelli}}]{2017ExA....44...25D}
{De Angelis}, A., {Tatischeff}, V., {Tavani}, M., {et~al.} 2017, Experimental
  Astronomy, 44, 25, \dodoi{10.1007/s10686-017-9533-6}

\bibitem[{{Dermer} \& {Menon}(2009)}]{2009herb.book.....D}
{Dermer}, C.~D., \& {Menon}, G. 2009, {High Energy Radiation from Black Holes:
  Gamma Rays, Cosmic Rays, and Neutrinos}

\bibitem[{{Dubner} {et~al.}(1998){Dubner}, {Holdaway}, {Goss}, \&
  {Mirabel}}]{1998AJ....116.1842D}
{Dubner}, G.~M., {Holdaway}, M., {Goss}, W.~M., \& {Mirabel}, I.~F. 1998, \aj,
  116, 1842, \dodoi{10.1086/300537}

\bibitem[{{Eikenberry} {et~al.}(2001){Eikenberry}, {Cameron}, {Fierce}, {Kull},
  {Dror}, {Houck}, \& {Margon}}]{2001ApJ...561.1027E}
{Eikenberry}, S.~S., {Cameron}, P.~B., {Fierce}, B.~W., {et~al.} 2001, \apj,
  561, 1027, \dodoi{10.1086/323380}

\bibitem[{{Fabrika}(2004)}]{2004ASPRv..12....1F}
{Fabrika}, S. 2004, \apspr, 12, 1.
\newblock \doarXiv{astro-ph/0603390}

\bibitem[{{Fang} {et~al.}(2020){Fang}, {Charles}, \&
  {Blandford}}]{2020ApJ...889L...5F}
{Fang}, K., {Charles}, E., \& {Blandford}, R.~D. 2020, \apjl, 889, L5,
  \dodoi{10.3847/2041-8213/ab62b8}

\bibitem[{{Finke} {et~al.}(2008){Finke}, {Dermer}, \&
  {B{\"o}ttcher}}]{2008ApJ...686..181F}
{Finke}, J.~D., {Dermer}, C.~D., \& {B{\"o}ttcher}, M. 2008, \apj, 686, 181,
  \dodoi{10.1086/590900}

\bibitem[{{Geldzahler} {et~al.}(1980){Geldzahler}, {Pauls}, \&
  {Salter}}]{1980A&A....84..237G}
{Geldzahler}, B.~J., {Pauls}, T., \& {Salter}, C.~J. 1980, \aap, 84, 237

\bibitem[{{Inoue} \& {Takahara}(1996)}]{1996ApJ...463..555I}
{Inoue}, S., \& {Takahara}, F. 1996, \apj, 463, 555, \dodoi{10.1086/177270}

\bibitem[{{Inoue} {et~al.}(2012){Inoue}, {Yamazaki}, {Inutsuka}, \&
  {Fukui}}]{IYI12a}
{Inoue}, T., {Yamazaki}, R., {Inutsuka}, S.-i., \& {Fukui}, Y. 2012, \apj, 744,
  71, \dodoi{10.1088/0004-637X/744/1/71}

\bibitem[{{Inoue} \& {Tanaka}(2016)}]{2016ApJ...828...13I}
{Inoue}, Y., \& {Tanaka}, Y.~T. 2016, \apj, 828, 13,
  \dodoi{10.3847/0004-637X/828/1/13}

\bibitem[{{Kafexhiu} {et~al.}(2014){Kafexhiu}, {Aharonian}, {Taylor}, \&
  {Vila}}]{2014PhRvD..90l3014K}
{Kafexhiu}, E., {Aharonian}, F., {Taylor}, A.~M., \& {Vila}, G.~S. 2014, \prd,
  90, 123014, \dodoi{10.1103/PhysRevD.90.123014}

\bibitem[{{Kelner} {et~al.}(2006){Kelner}, {Aharonian}, \& {Bugayov}}]{kab06}
{Kelner}, S.~R., {Aharonian}, F.~A., \& {Bugayov}, V.~V. 2006, \prd, 74,
  034018, \dodoi{10.1103/PhysRevD.74.034018}

\bibitem[{{Kimura} {et~al.}(2015){Kimura}, {Murase}, \& {Toma}}]{kmt15}
{Kimura}, S.~S., {Murase}, K., \& {Toma}, K. 2015, \apj, 806, 159,
  \dodoi{10.1088/0004-637X/806/2/159}

\bibitem[{{Kimura} {et~al.}(2018){Kimura}, {Murase}, \& {Zhang}}]{KMZ18a}
{Kimura}, S.~S., {Murase}, K., \& {Zhang}, B.~T. 2018, \prd, 97, 023026,
  \dodoi{10.1103/PhysRevD.97.023026}

\bibitem[{{Konigl}(1983)}]{1983MNRAS.205..471K}
{Konigl}, A. 1983, \mnras, 205, 471, \dodoi{10.1093/mnras/205.2.471}

\bibitem[{{MAGIC Collaboration} {et~al.}(2018){MAGIC Collaboration}, {Ahnen},
  {Ansoldi}, {Antonelli}, {Arcaro}, {Babi{\'c}}, {Banerjee}, {Bangale}, {Barres
  de Almeida}, {Barrio}, {Becerra Gonz{\'a}lez}, {Bednarek}, {Bernardini},
  {Berti}, {Biasuzzi}, {Biland}, {Blanch}, {Bonnefoy}, {Bonnoli}, {Borracci},
  {Carosi}, {Carosi}, {Chatterjee}, {Colin}, {Colombo}, {Contreras}, {Cortina},
  {Covino}, {Cumani}, {da Vela}, {Dazzi}, {de Angelis}, {de Lotto}, {de O{\~n}a
  Wilhelmi}, {di Pierro}, {Doert}, {Dom{\'\i}nguez}, {Dominis Prester},
  {Dorner}, {Doro}, {Einecke}, {Eisenacher Glawion}, {Elsaesser},
  {Engelkemeier}, {Fallah Ramazani}, {Fern{\'a}ndez-Barral}, {Fidalgo},
  {Fonseca}, {Font}, {Fruck}, {Galindo}, {Garc{\'\i}a L{\'o}pez},
  {Garczarczyk}, {Gaug}, {Giammaria}, {Godinovi{\'c}}, {Gora}, {Griffiths},
  {Guberman}, {Hadasch}, {Hahn}, {Hassan}, {Hayashida}, {Herrera}, {Hose},
  {Hrupec}, {Hughes}, {Ishio}, {Konno}, {Kubo}, {Kushida},
  {Kuve{\v{z}}di{\'c}}, {Lelas}, {Lindfors}, {Lombardi}, {Longo}, {L{\'o}pez},
  {L{\'o}pez-Oramas}, {Majumdar}, {Makariev}, {Maneva}, {Manganaro},
  {Mannheim}, {Maraschi}, {Mariotti}, {Mart{\'\i}nez}, {Mazin}, {Menzel},
  {Minev}, {Mirzoyan}, {Moralejo}, {Moreno}, {Moretti}, {Munar-Adrover},
  {Neustroev}, {Niedzwiecki}, {Nievas Rosillo}, {Nilsson}, {Nishijima}, {Noda},
  {Nogu{\'e}s}, {Paiano}, {Palacio}, {Paneque}, {Paoletti}, {Paredes},
  {Paredes-Fortuny}, {Pedaletti}, {Peresano}, {Perri}, {Persic}, {Prada
  Moroni}, {Prand ini}, {Puljak}, {Garcia}, {Reichardt}, {Rhode}, {Rib{\'o}},
  {Rico}, {Saito}, {Satalecka}, {Schroeder}, {Schweizer}, {Shore},
  {Sillanp{\"a}{\"a}}, {Sitarek}, {{\v{S}}nidari{\'c}}, {Sobczynska},
  {Stamerra}, {Strzys}, {Suri{\'c}}, {Takalo}, {Tavecchio}, {Temnikov},
  {Terzi{\'c}}, {Tescaro}, {Teshima}, {Torres}, {Torres-Alb{\`a}}, {Treves},
  {Vanzo}, {Vazquez Acosta}, {Vovk}, {Ward}, {Will}, {Wu}, {Zari{\'c}},
  {H.~E.~S.~S. Collaboration}, {Abdalla}, {Abramowski}, {Aharonian}, {Ait
  Benkhali}, {Akhperjanian}, {Andersson}, {Ang{\"u}ner}, {Arakawa}, {Arrieta},
  {Aubert}, {Backes}, {Balzer}, {Barnard}, {Becherini}, {Becker Tjus}, {Berge},
  {Bernhard}, {Bernl{\"o}hr}, {Blackwell}, {B{\"o}ttcher}, {Boisson},
  {Bolmont}, {Bordas}, {Bregeon}, {Brun}, {Brun}, {Bryan}, {B{\"u}chele},
  {Bulik}, {Capasso}, {Carr}, {Casanova}, {Cerruti}, {Chakraborty},
  {Chalme-Calvet}, {Chaves}, {Chen}, {Chevalier}, {Chr{\'e}tien}, {Coffaro},
  {Colafrancesco}, {Cologna}, {Condon}, {Conrad}, {Cui}, {Davids}, {Decock},
  {Degrange}, {Deil}, {Devin}, {Dewilt}, {Dirson}, {Djannati-Ata{\"\i}},
  {Domainko}, {Donath}, {Drury}, {Dutson}, {Dyks}, {Edwards}, {Egberts},
  {Eger}, {Ernenwein}, {Eschbach}, {Farnier}, {Fegan}, {Fernandes}, {Fiasson},
  {Fontaine}, {F{\"o}rster}, {Funk}, {F{\"u}{\ss}ling}, {Gabici}, {Gajdus},
  {Gallant}, {Garrigoux}, {Giavitto}, {Giebels}, {Glicenstein}, {Gottschall},
  {Goyal}, {Grondin}, {Hahn}, {Haupt}, {Hawkes}, {Heinzelmann}, {Henri},
  {Hermann}, {Hervet}, {Hinton}, {Hofmann}, {Hoischen}, {Holler}, {Horns},
  {Ivascenko}, {Iwasaki}, {Jacholkowska}, {Jamrozy}, {Janiak}, {Jankowsky},
  {Jankowsky}, {Jingo}, {Jogler}, {Jouvin}, {Jung-Richardt}, {Kastendieck},
  {Katarzy{\'n}ski}, {Katsuragawa}, {Katz}, {Kerszberg}, {Khangulyan},
  {Kh{\'e}lifi}, {Kieffer}, {King}, {Klepser}, {Klochkov}, {Klu{\'z}niak},
  {Kolitzus}, {Komin}, {Kosack}, {Krakau}, {Kraus}, {Kr{\"u}ger}, {Laffon},
  {Lamanna}, {Lau}, {Lees}, {Lefaucheur}, {Lefranc}, {Lemi{\`e}re},
  {Lemoine-Goumard}, {Lenain}, {Leser}, {Lohse}, {Lorentz}, {Liu},
  {L{\'o}pez-Coto}, {Lypova}, {Marandon}, {Marcowith}, {Mariaud}, {Marx},
  {Maurin}, {Maxted}, {Mayer}, {Meintjes}, {Meyer}, {Mitchell}, {Moderski},
  {Mohamed}, {Mohrmann}, {Mor{\r{a}}}, {Moulin}, {Murach}, {Nakashima}, {de
  Naurois}, {Niederwanger}, {Niemiec}, {Oakes}, {O'Brien}, {Odaka}, {{\"O}ttl},
  {Ohm}, {Ostrowski}, {Oya}, {Padovani}, {Panter}, {Parsons}, {Pekeur},
  {Pelletier}, {Perennes}, {Petrucci}, {Peyaud}, {Piel}, {Pita}, {Poon},
  {Prokhorov}, {Prokoph}, {P{\"u}hlhofer}, {Punch}, {Quirrenbach}, {Raab},
  {Reimer}, {Reimer}, {Renaud}, {de Los Reyes}, {Richter}, {Rieger}, {Romoli},
  {Rowell}, {Rudak}, {Rulten}, {Safi-Harb}, {Sahakian}, {Saito}, {Salek},
  {Sanchez}, {Santangelo}, {Sasaki}, {Schlickeiser}, {Sch{\"u}ssler}, {Schulz},
  {Schwanke}, {Schwemmer}, {Seglar-Arroyo}, {Settimo}, {Seyffert}, {Shafi},
  {Shilon}, {Simoni}, {Sol}, {Spanier}, {Spengler}, {Spies}, {Stawarz},
  {Steenkamp}, {Stegmann}, {Stycz}, {Sushch}, {Takahashi}, {Tavernet},
  {Tavernier}, {Taylor}, {Terrier}, {Tibaldo}, {Tiziani}, {Tluczykont},
  {Trichard}, {Tsuji}, {Tuffs}, {Uchiyama}, {van der Walt}, {van Eldik}, {van
  Rensburg}, {van Soelen}, {Vasileiadis}, {Veh}, {Venter}, {Viana}, {Vincent},
  {Vink}, {Voisin}, {V{\"o}lk}, {Vuillaume}, {Wadiasingh}, {Wagner}, {Wagner},
  {Wagner}, {White}, {Wierzcholska}, {Willmann}, {W{\"o}rnlein}, {Wouters},
  {Yang}, {Zabalza}, {Zaborov}, {Zacharias}, {Zanin}, {Zdziarski}, {Zech},
  {Zefi}, {Ziegler}, \& {Zywucka}}]{2018A&A...612A..14M}
{MAGIC Collaboration}, {Ahnen}, M.~L., {Ansoldi}, S., {et~al.} 2018, \aap, 612,
  A14, \dodoi{10.1051/0004-6361/201731169}

\bibitem[{{Marshall} {et~al.}(2002){Marshall}, {Canizares}, \&
  {Schulz}}]{2002ApJ...564..941M}
{Marshall}, H.~L., {Canizares}, C.~R., \& {Schulz}, N.~S. 2002, \apj, 564, 941,
  \dodoi{10.1086/324398}

\bibitem[{{Meyer} {et~al.}(2017){Meyer}, {Sparks}, {Georganopoulos}, {van der
  Marel}, {Anderson}, {Sohn}, {Biretta}, {Norman}, {Chiaberge}, \&
  {Perlman}}]{2017Galax...5....8M}
{Meyer}, E., {Sparks}, W., {Georganopoulos}, M., {et~al.} 2017, Galaxies, 5, 8,
  \dodoi{10.3390/galaxies5010008}

\bibitem[{{Meyer} \& {Georganopoulos}(2014)}]{2014ApJ...780L..27M}
{Meyer}, E.~T., \& {Georganopoulos}, M. 2014, \apjl, 780, L27,
  \dodoi{10.1088/2041-8205/780/2/L27}

\bibitem[{{Moiseev} \& {Amego Team}(2017)}]{2017ICRC...35..798M}
{Moiseev}, A., \& {Amego Team}. 2017, International Cosmic Ray Conference, 301,
  798

\bibitem[{{Mullin} \& {Hardcastle}(2009)}]{2009MNRAS.398.1989M}
{Mullin}, L.~M., \& {Hardcastle}, M.~J. 2009, \mnras, 398, 1989,
  \dodoi{10.1111/j.1365-2966.2009.15232.x}

\bibitem[{{Murase} {et~al.}(2013){Murase}, {Ahlers}, \& {Lacki}}]{mal13}
{Murase}, K., {Ahlers}, M., \& {Lacki}, B.~C. 2013, \prd, 88, 121301,
  \dodoi{10.1103/PhysRevD.88.121301}

\bibitem[{{Murase} {et~al.}(2020){Murase}, {Kimura}, \&
  {M{\'e}sz{\'a}ros}}]{2020PhRvL.125a1101M}
{Murase}, K., {Kimura}, S.~S., \& {M{\'e}sz{\'a}ros}, P. 2020, \prl, 125,
  011101, \dodoi{10.1103/PhysRevLett.125.011101}

\bibitem[{{Panferov}(2017)}]{2017A&A...599A..77P}
{Panferov}, A.~A. 2017, \aap, 599, A77, \dodoi{10.1051/0004-6361/201629256}

\bibitem[{{Park} {et~al.}(2015){Park}, {Caprioli}, \& {Spitkovsky}}]{PCS15a}
{Park}, J., {Caprioli}, D., \& {Spitkovsky}, A. 2015, Physical Review Letters,
  114, 085003, \dodoi{10.1103/PhysRevLett.114.085003}

\bibitem[{{Park} {et~al.}(2019){Park}, {Hada}, {Kino}, {Nakamura}, {Hodgson},
  {Ro}, {Cui}, {Asada}, {Algaba}, {Sawada-Satoh}, {Lee}, {Cho}, {Shen},
  {Jiang}, {Trippe}, {Niinuma}, {Sohn}, {Jung}, {Zhao}, {Wajima}, {Tazaki},
  {Honma}, {An}, {Akiyama}, {Byun}, {Kim}, {Zhang}, {Cheng}, {Kobayashi},
  {Shibata}, {Lee}, {Roh}, {Oh}, {Yeom}, {Jung}, {Oh}, {Kim}, {Hwang}, \&
  {Hagiwara}}]{2019ApJ...887..147P}
{Park}, J., {Hada}, K., {Kino}, M., {et~al.} 2019, \apj, 887, 147,
  \dodoi{10.3847/1538-4357/ab5584}

\bibitem[{{Rasul} {et~al.}(2019){Rasul}, {Chadwick}, {Graham}, \&
  {Brown}}]{2019MNRAS.485.2970R}
{Rasul}, K., {Chadwick}, P.~M., {Graham}, J.~A., \& {Brown}, A.~M. 2019,
  \mnras, 485, 2970, \dodoi{10.1093/mnras/stz559}

\bibitem[{{Reynolds}(2008)}]{2008ARA&A..46...89R}
{Reynolds}, S.~P. 2008, \araa, 46, 89,
  \dodoi{10.1146/annurev.astro.46.060407.145237}

\bibitem[{{Reynoso} \& {Carulli}(2019)}]{2019APh...109...25R}
{Reynoso}, M.~M., \& {Carulli}, A.~M. 2019, Astroparticle Physics, 109, 25,
  \dodoi{10.1016/j.astropartphys.2019.02.003}

\bibitem[{{Safi-Harb} \& {{\"O}gelman}(1997)}]{1997ApJ...483..868S}
{Safi-Harb}, S., \& {{\"O}gelman}, H. 1997, \apj, 483, 868,
  \dodoi{10.1086/304274}

\bibitem[{{Safi-Harb} \& {Petre}(1999)}]{1999ApJ...512..784S}
{Safi-Harb}, S., \& {Petre}, R. 1999, \apj, 512, 784, \dodoi{10.1086/306803}

\bibitem[{{Sambruna} {et~al.}(2004){Sambruna}, {Gambill}, {Maraschi},
  {Tavecchio}, {Cerutti}, {Cheung}, {Urry}, \& {Chartas}}]{2004ApJ...608..698S}
{Sambruna}, R.~M., {Gambill}, J.~K., {Maraschi}, L., {et~al.} 2004, \apj, 608,
  698, \dodoi{10.1086/383124}

\bibitem[{{Santana} {et~al.}(2014){Santana}, {Barniol Duran}, \&
  {Kumar}}]{2014ApJ...785...29S}
{Santana}, R., {Barniol Duran}, R., \& {Kumar}, P. 2014, \apj, 785, 29,
  \dodoi{10.1088/0004-637X/785/1/29}

\bibitem[{{Seward} {et~al.}(1980){Seward}, {Grindlay}, {Seaquist}, \&
  {Gilmore}}]{1980Natur.287..806S}
{Seward}, F., {Grindlay}, J., {Seaquist}, E., \& {Gilmore}, W. 1980, \nat, 287,
  806, \dodoi{10.1038/287806a0}

\bibitem[{{Skilling}(1975)}]{1975MNRAS.173..255S}
{Skilling}, J. 1975, \mnras, 173, 255, \dodoi{10.1093/mnras/173.2.255}

\bibitem[{{Stawarz} \& {Petrosian}(2008)}]{sp08}
{Stawarz}, {\L}., \& {Petrosian}, V. 2008, \apj, 681, 1725,
  \dodoi{10.1086/588813}

\bibitem[{{Strong} {et~al.}(2007){Strong}, {Moskalenko}, \& {Ptuskin}}]{SMP07a}
{Strong}, A.~W., {Moskalenko}, I.~V., \& {Ptuskin}, V.~S. 2007, Annual Review
  of Nuclear and Particle Science, 57, 285,
  \dodoi{10.1146/annurev.nucl.57.090506.123011}

\bibitem[{{Su} {et~al.}(2018){Su}, {Zhou}, {Yang}, {Chen}, {Chen}, \&
  {Zhang}}]{2018ApJ...863..103S}
{Su}, Y., {Zhou}, X., {Yang}, J., {et~al.} 2018, \apj, 863, 103,
  \dodoi{10.3847/1538-4357/aad04e}

\bibitem[{{Sudoh} {et~al.}(2020){Sudoh}, {Inoue}, \&
  {Khangulyan}}]{2020ApJ...889..146S}
{Sudoh}, T., {Inoue}, Y., \& {Khangulyan}, D. 2020, \apj, 889, 146,
  \dodoi{10.3847/1538-4357/ab6442}

\bibitem[{{Sun} {et~al.}(2019){Sun}, {Yang}, {Liu}, {Xi}, \&
  {Wang}}]{2019A&A...626A.113S}
{Sun}, X.-N., {Yang}, R.-Z., {Liu}, B., {Xi}, S.-Q., \& {Wang}, X.-Y. 2019,
  \aap, 626, A113, \dodoi{10.1051/0004-6361/201935621}

\bibitem[{{Takahara}(1990)}]{Tak90a}
{Takahara}, F. 1990, Progress of Theoretical Physics, 83, 1071,
  \dodoi{10.1143/PTP.83.1071}

\bibitem[{{Tanaka} \& {Takahara}(2011)}]{2011ApJ...741...40T}
{Tanaka}, S.~J., \& {Takahara}, F. 2011, \apj, 741, 40,
  \dodoi{10.1088/0004-637X/741/1/40}

\bibitem[{{Tanaka} \& {Takahara}(2013)}]{2013MNRAS.429.2945T}
---. 2013, \mnras, 429, 2945, \dodoi{10.1093/mnras/sts528}

\bibitem[{{Tavecchio} {et~al.}(2000){Tavecchio}, {Maraschi}, {Sambruna}, \&
  {Urry}}]{TMS00a}
{Tavecchio}, F., {Maraschi}, L., {Sambruna}, R.~M., \& {Urry}, C.~M. 2000,
  \apjl, 544, L23, \dodoi{10.1086/317292}

\bibitem[{{The Fermi-LAT collaboration}(2019)}]{2019arXiv190210045T}
{The Fermi-LAT collaboration}. 2019, arXiv e-prints, arXiv:1902.10045.
\newblock \doarXiv{1902.10045}

\bibitem[{{The IceCube-Gen2 Collaboration} {et~al.}(2020){The IceCube-Gen2
  Collaboration}, {:}, {Aartsen}, {Abbasi}, {Ackermann}, {Adams}, {Aguilar},
  {Ahlers}, {Ahrens}, {Alispach}, {Allison}, {Amin}, {Andeen}, {Anderson},
  {Ansseau}, {Anton}, {Arg{\"u}elles}, {Arlen}, {Auffenberg}, {Axani},
  {Bagherpour}, {Bai}, {Balagopal V.}, {Barbano}, {Bartos}, {Bastian}, {Basu},
  {Baum}, {Baur}, {Bay}, {Beatty}, {Becker}, {Becker Tjus}, {BenZvi}, {Berley},
  {Bernardini}, {Besson}, {Binder}, {Bindig}, {Blaufuss}, {Blot}, {Bohm},
  {Bohmer}, {B{\"o}ser}, {Botner}, {B{\"o}ttcher}, {Bourbeau}, {Bourbeau},
  {Bradascio}, {Braun}, {Bron}, {Brostean-Kaiser}, {Burgman}, {Burley},
  {Buscher}, {Busse}, {Bustamante}, {Campana}, {Carnie-Bronca}, {Carver},
  {Chen}, {Chen}, {Cheung}, {Chirkin}, {Choi}, {Clark}, {Clark}, {Classen},
  {Coleman}, {Collin}, {Connolly}, {Conrad}, {Coppin}, {Correa}, {Cowen},
  {Cross}, {Dave}, {Deaconu}, {De Clercq}, {DeLaunay}, {De Kockere},
  {Dembinski}, {Deoskar}, {De Ridder}, {Desai}, {Desiati}, {de Vries}, {de
  Wasseige}, {de With}, {DeYoung}, {Dharani}, {Diaz}, {D{\'\i}az-V{\'e}lez},
  {Dujmovic}, {Dunkman}, {DuVernois}, {Dvorak}, {Ehrhardt}, {Eller}, {Engel},
  {Evans}, {Evenson}, {Fahey}, {Farrag}, {Fazely}, {Felde}, {Fienberg},
  {Filimonov}, {Finley}, {Fischer}, {Fox}, {Franckowiak}, {Friedman}, {Fritz},
  {Gaisser}, {Gallagher}, {Ganster}, {Garcia-Fernand ez}, {Garrappa},
  {Gartner}, {Gerhardt}, {Gernhaeuser}, {Ghadimi}, {Glaser}, {Glauch},
  {Gl{\"u}senkamp}, {Goldschmidt}, {Gonzalez}, {Goswami}, {Grant},
  {Gr{\'e}goire}, {Griffith}, {Griswold}, {G{\"u}nd{\"u}z}, {Haack},
  {Hallgren}, {Halliday}, {Halve}, {Halzen}, {Hanson}, {Hanson}, {Hardin},
  {Haugen}, {Haungs}, {Hauser}, {Hebecker}, {Heinen}, {Heix}, {Helbing},
  {Hellauer}, {Henningsen}, {Hickford}, {Hignight}, {Hill}, {Hill}, {Hoffman},
  {Hoffmann}, {Hoffmann}, {Hoinka}, {Hokanson-Fasig}, {Holzapfel}, {Hoshina},
  {Huang}, {Huber}, {Huber}, {Huege}, {Hughes}, {Hultqvist}, {H{\"u}nnefeld},
  {Hussain}, {In}, {Iovine}, {Ishihara}, {Jansson}, {Japaridze}, {Jeong},
  {Jones}, {Jonske}, {Joppe}, {Kalekin}, {Kang}, {Kang}, {Kang}, {Kappes},
  {Kappesser}, {Karg}, {Karl}, {Karle}, {Katori}, {Katz}, {Kauer}, {Keivani},
  {Kellermann}, {Kelley}, {Kheirand ish}, {Kim}, {Kin}, {Kintscher}, {Kiryluk},
  {Kittler}, {Kleifges}, {Klein}, {Koirala}, {Kolanoski}, {K{\"o}pke},
  {Kopper}, {Kopper}, {Koskinen}, {Koundal}, {Kovacevich}, {Kowalski},
  {Krauss}, {Krings}, {Kr{\"u}ckl}, {Kulacz}, {Kurahashi}, {Lagunas Gualda},
  {Lahmann}, {Lanfranchi}, {Larson}, {Latif}, {Lauber}, {Lazar}, {Leonard},
  {Leszczy{\'n}ska}, {Li}, {Liu}, {Lohfink}, {LoSecco}, {Lozano Mariscal},
  {Lu}, {Lucarelli}, {Ludwig}, {L{\"u}nemann}, {Luszczak}, {Lyu}, {Ma},
  {Madsen}, {Maggi}, {Mahn}, {Makino}, {Mallik}, {Mancina}, {Mandalia},
  {Mari{\textcommabelow s}}, {Marka}, {Marka}, {Maruyama}, {Mase}, {Maunu},
  {McNally}, {Meagher}, {Medina}, {Meier}, {Meighen-Berger}, {Merz}, {Meyers},
  {Micallef}, {Mockler}, {Moment{\'e}}, {Montaruli}, {Moore}, {Morse},
  {Moulai}, {Muth}, {Naab}, {Nagai}, {Nam}, {Naumann}, {Necker}, {Neer},
  {Nelles}, {Nguy{\^e}n}, {Niederhausen}, {Nisa}, {Nowicki}, {Nygren},
  {Oberla}, {Obertacke Pollmann}, {Oehler}, {Olivas}, {O'Sullivan}, {Pan},
  {Pand ya}, {Pankova}, {Papp}, {Park}, {Parker}, {Paudel}, {Peiffer},
  {P{\'e}rez de los Heros}, {Petersen}, {Philippen}, {Pieloth}, {Pieper},
  {Pinfold}, {Pizzuto}, {Plaisier}, {Plum}, {Popovych}, {Porcelli}, {Prado
  Rodriguez}, {Price}, {Przybylski}, {Raab}, {Raissi}, {Rameez}, {Rauch},
  {Rawlins}, {Rea}, {Rehman}, {Reimann}, {Renschler}, {Renzi}, {Resconi},
  {Reusch}, {Rhode}, {Richman}, {Riedel}, {Riegel}, {Roberts}, {Robertson},
  {Roellinghoff}, {Rongen}, {Rott}, {Ruhe}, {Ryckbosch}, {Rysewyk Cantu},
  {Safa}, {Sanchez Herrera}, {Sand rock}, {Sandroos}, {Sandstrom}, {Santander},
  {Sarkar}, {Sarkar}, {Satalecka}, {Scharf}, {Schaufel}, {Schieler},
  {Schlunder}, {Schmidt}, {Schneider}, {Schneider}, {Schr{\"o}der},
  {Schumacher}, {Sclafani}, {Seckel}, {Seunarine}, {Shaevitz}, {Sharma},
  {Shefali}, {Silva}, {Smith}, {Smithers}, {Snihur}, {Soedingrekso}, {Soldin},
  {S{\"o}ldner-Rembold}, {Song}, {Southall}, {Spiczak}, {Spiering},
  {Stachurska}, {Stamatikos}, {Stanev}, {Stein}, {Stettner}, {Steuer},
  {Stezelberger}, {Stokstad}, {Strotjohann}, {St{\"u}rwald}, {Stuttard},
  {Sullivan}, {Taboada}, {Taketa}, {Tanaka}, {Tenholt}, {Ter-Antonyan},
  {Terliuk}, {Tilav}, {Tollefson}, {Tomankova}, {T{\"o}nnis}, {Torres},
  {Toscano}, {Tosi}, {Trettin}, {Tselengidou}, {Tung}, {Turcati}, {Turcotte},
  {Turley}, {Twagirayezu}, {Ty}, {Unger}, {Unland Elorrieta}, {Vand enbroucke},
  {van Eijk}, {van Eijndhoven}, {Vannerom}, {van Santen}, {Veberic},
  {Verpoest}, {Vieregg}, {Vraeghe}, {Walck}, {Watson}, {Weaver}, {Weindl},
  {Weinstock}, {Weiss}, {Weldert}, {Welling}, {Wendt}, {Werthebach},
  {Whitehorn}, {Wiebe}, {Wiebusch}, {Williams}, {Wissel}, {Wolf}, {Wood},
  {Woschnagg}, {Wrede}, {Wren}, {Wulff}, {Xu}, {Xu}, {Yanez}, {Yoshida},
  {Yuan}, {Zhang}, {Zierke}, \& {Z{\"o}cklein}}]{2020arXiv200804323T}
{The IceCube-Gen2 Collaboration}, {:}, {Aartsen}, M.~G., {et~al.} 2020, arXiv
  e-prints, arXiv:2008.04323.
\newblock \doarXiv{2008.04323}

\bibitem[{{Tomita} {et~al.}(2019){Tomita}, {Ohira}, \&
  {Yamazaki}}]{2019ApJ...886...54T}
{Tomita}, S., {Ohira}, Y., \& {Yamazaki}, R. 2019, \apj, 886, 54,
  \dodoi{10.3847/1538-4357/ab4a10}

\bibitem[{{van den Heuvel}(1981)}]{1981VA.....25...95V}
{van den Heuvel}, E.~P.~J. 1981, Vistas in Astronomy, 25, 95,
  \dodoi{10.1016/0083-6656(81)90050-7}

\bibitem[{{Wardle} \& {Aaron}(1997)}]{1997MNRAS.286..425W}
{Wardle}, J.~F.~C., \& {Aaron}, S.~E. 1997, \mnras, 286, 425,
  \dodoi{10.1093/mnras/286.2.425}

\bibitem[{{Watson} {et~al.}(1983){Watson}, {Willingale}, {Grindlay}, \&
  {Seward}}]{1983ApJ...273..688W}
{Watson}, M.~G., {Willingale}, R., {Grindlay}, J.~E., \& {Seward}, F.~D. 1983,
  \apj, 273, 688, \dodoi{10.1086/161403}

\bibitem[{{Werner} {et~al.}(2012){Werner}, {Murphy}, {Livingston}, {Gorjian},
  {Jones}, {Meier}, \& {Lawrence}}]{2012ApJ...759...86W}
{Werner}, M.~W., {Murphy}, D.~W., {Livingston}, J.~H., {et~al.} 2012, \apj,
  759, 86, \dodoi{10.1088/0004-637X/759/2/86}

\bibitem[{{Xing} {et~al.}(2019){Xing}, {Wang}, {Zhang}, {Chen}, \&
  {Jithesh}}]{2019ApJ...872...25X}
{Xing}, Y., {Wang}, Z., {Zhang}, X., {Chen}, Y., \& {Jithesh}, V. 2019, \apj,
  872, 25, \dodoi{10.3847/1538-4357/aafc60}

\bibitem[{{Yamamoto} {et~al.}(2008){Yamamoto}, {Ito}, {Ishigami}, {Fujishita},
  {Kawase}, {Kawamura}, {Mizuno}, {Onishi}, {Mizuno}, {McClure-Griffiths}, \&
  {Fukui}}]{2008PASJ...60..715Y}
{Yamamoto}, H., {Ito}, S., {Ishigami}, S., {et~al.} 2008, \pasj, 60, 715,
  \dodoi{10.1093/pasj/60.4.715}

\bibitem[{{Yamauchi} {et~al.}(1994){Yamauchi}, {Kawai}, \&
  {Aoki}}]{1994PASJ...46L.109Y}
{Yamauchi}, S., {Kawai}, N., \& {Aoki}, T. 1994, \pasj, 46, L109

\bibitem[{{Zealey} {et~al.}(1980){Zealey}, {Dopita}, \&
  {Malin}}]{1980MNRAS.192..731Z}
{Zealey}, W.~J., {Dopita}, M.~A., \& {Malin}, D.~F. 1980, \mnras, 192, 731,
  \dodoi{10.1093/mnras/192.4.731}

\bibitem[{{Zhang} {et~al.}(2018){Zhang}, {Du}, {Guo}, {Zhang}, {Chen}, {Liang},
  \& {Zhang}}]{2018ApJ...858...27Z}
{Zhang}, J., {Du}, S.-s., {Guo}, S.-C., {et~al.} 2018, \apj, 858, 27,
  \dodoi{10.3847/1538-4357/aab9b2}

\end{thebibliography}


\end{document}